\documentclass[preprint,12pt]{elsarticle}
\usepackage{graphicx}
\usepackage{amsfonts}
\usepackage{amsmath}
\usepackage{url}

\def\var{{\rm var}}
\def\PNL{{\mathcal P}}  
\def\dPNL{\delta\PNL}  
\def\E{{\bf E}}
\def\A{{\bf A}}
\def\S{{\bf S}}
\def\C{{\bf C}}
\def\Sh{{\mathcal S}}
\def\ve{\varepsilon}
\def\T{{\dagger}}

\journal{Physica A}

\begin{document}

\begin{frontmatter}
\title{Optimal Allocation of Trend Following Strategies}

\author{Denis~S.~Grebenkov}
\address{
Laboratoire de Physique de la Mati\`{e}re Condens\'{e}e, \\ 
CNRS -- Ecole Polytechnique, 91128 Palaiseau, France}
\ead{denis.grebenkov@polytechnique.edu}

\author{Jeremy~Serror}
\address{
John Locke Investment, 38 Avenue Franklin Roosevelt, 77210 Fontainebleau-Avon, France}
  \ead{jeremy.serror@gmail.com}

\date{\today}

\begin{abstract}
We consider a portfolio allocation problem for trend following (TF)
strategies on multiple correlated assets.  Under simplifying
assumptions of a Gaussian market and linear TF strategies, we derive
analytical formulas for the mean and variance of the portfolio return.
We construct then the optimal portfolio that maximizes risk-adjusted
return by accounting for inter-asset correlations.  The dynamic
allocation problem for $n$ assets is shown to be equivalent to the
classical static allocation problem for $n^2$ virtual assets that
include lead-lag corrections in positions of TF strategies.  The
respective roles of asset auto-correlations and inter-asset
correlations are investigated in depth for the two-asset case and a
sector model.  In contrast to the principle of diversification
suggesting to treat uncorrelated assets, we show that inter-asset
correlations allow one to estimate apparent trends more reliably and
to adjust the TF positions more efficiently.  If properly accounted
for, inter-asset correlations are not deteriorative but beneficial for
portfolio management that can open new profit opportunities for trend
followers.
\end{abstract}



\end{frontmatter}

\section{Introduction}

For decades, market participants have attempted to detect potential
trends in asset price fluctuations on exchange markets.  In systematic
trading, trend following (TF) strategies that generate buy or sell
signals to adjust their market exposure according to past price
variations, were developed to profit from trends at various time
horizons \cite{Covel,Clenow,Fung01,Asness13}.  While its actual
profitability is highly controversial
\cite{Potters05,Chan96,Jegadeesh01,Moskowitz12}, trend following
remains a widely used strategy among professional asset managers.
Since many traders search for the same profit opportunities, the
expected (net) gains are small, especially at short times, and are
subject to stochastic fluctuations.  In order to enhance profit and
reduce risk, fund managers build diversified portfolios, aiming to
decorrelate constituent TF strategies as much as possible.  Our goal
is to show that this conventional approach leads to suboptimal
portfolios.  In particular, we illustrate that inter-asset
correlations, if accounted for properly, facilitate trend detection
and thus significantly improve the risk-adjusted portfolio return.

In a previous work, we considered a linear TF strategy applied to an
asset with auto-correlated returns \cite{Grebenkov14}.  This model
relies on the ability of market participants to assess market
auto-correlation or, equivalently, the excess variance.  An explicit
persistence in asset returns was introduced in order to study trend
following from a risk-reward perspective.  Modeling price persistence
by adding a stochastic trend term to a Gaussian market model, we
derived analytical formulas for the mean and variance of the strategy
profit-and-losses (P\&L).  Given market transaction costs, we were
able, for instance, to compute a threshold in auto-correlation below
which trend follower has no hope to realize profit in real market
conditions.  Fund managers often use such criteria to select a set of
assets/markets that are of interest for trend following trading.  Many
examples of TF strategies applied to stock markets, foreign exchange
markets, and commodities were reported
\cite{Clenow,Allen90,Wilcox05,Szakmary10}.  In the financial industry,
diversified funds apply TF strategies to a large number of assets with
hope to benefit from the so-called diversification effect
\cite{Sharpe,Ilmanen12}.

In this paper we extend the model from \cite{Grebenkov14} to the
multivariate case, in which explicit stochastic trends and an
inter-asset covariance structure are introduced.  In particular, we
study how correlations in the market trends affect the portfolio
risk-reward profile.  In other words, while asset returns may exhibit
a given covariance structure, their trend component may have a
different one.  We aim at solving a portfolio optimization problem
taking into account the trend following nature of the trading
strategies.  Our goal is to show that failure to account for trend
correlations (i.e., only using asset returns covariance) leads to
suboptimal risk-adjusted portfolio return.

Starting from the seminal work of Markowitz \cite{Markowitz52}, modern
portfolio theory \cite{Merton71,Elton97,Pfaff} brought numerous
optimization techniques to the asset allocation problem
\cite{Campbell}.  The initial problem Markowitz considered was to find
portfolio weights, i.e., amount of capital allocated to each asset,
maximizing a portfolio mean-variance objective given expected market
returns and covariance structure.  The Markowitz model relies on the
ability of market participants to assess expected return and provides
a way to incorporate asset covariance into the investment process.  In
our approach, the asset expected return is substituted by the expected
excess variance (or auto-correlation) that characterizes market trends
\cite{Grebenkov14,Bouchaud98}.  We solve the problem of static
allocation of dynamic strategies by specifying the correlation
structure of trend and noise components of asset price fluctuations.
As another approach to the dynamic allocation problem, one often
considers a sequence of static portfolios in the so-called
multi-period Markowitz framework \cite{Garleanu13,Li98}.

We choose simple modeling assumptions from \cite{Grebenkov14} in an
effort to derive an exact solution to the problem:

(i) applying a TF strategy implicitly assumes persistence in price
variations.  We model price variations as a stochastic trend plus a
white noise.

(ii) real markets exhibit inter-asset (or cross) correlations
\cite{Embrechts,Fenn11}.  We introduce separately the correlation in
trends and the correlation in noises.  For instance, two assets can
exhibit similar long-term trends and be negatively correlated on the
short term.

Under these assumptions, we find that the static allocation problem in
which an optimal weight is assigned to each asset, leads to suboptimal
risk-adjusted return.  Even if the correlations in trend and noise are
equal, the application of a classical Markowitz approach to TF
strategies is suboptimal.  We then formulate a dynamic allocation
framework that leads to an improved risk-adjusted return of the
portfolio.  Our approach to dynamic allocation consists in correcting
each strategy signal by a linear combination of other strategy
signals.  This cross-correcting term can be seen as a lead-lag
correction \cite{Stoll90,Chan92}.  We show that the allocation problem
for $n$ dynamic strategies can be reduced to solving a static
Markowitz problem for a set of $n^2$ virtual assets with explicitly
derived expected returns and covariance structure.  We deduce the
simple rule of thumb: for two assets $i$ and $j$, given their
respective strategy signals $S^i_t$ and $S^j_t$, and cross-correlation
$\rho_{i,j}$ between $i$ and $j$, one should adjust the exposure of
the $i$-th signal proportionally to $-S^j_t
\rho_{i,j}$ and the exposure of the $j$-th signal proportionally to
$-S^i_t \rho_{i,j}$.  For instance, if both signals are positive, then
an increase in cross-correlation reduces exposure though the
cross-correcting term.  As the cross-correction for the $i$-th asset
is directly proportional to a linear combination of the $j$-th asset
past returns, we refer to it as a lead-lag term \cite{Stoll90,Chan92}.

The paper is organized as follows.  In Sec. \ref{sec:general}, we
introduce the standard mathematical tools to solve the portfolio
allocation problem.  In Sec. \ref{sec:two}, we study in detail the
two-asset portfolio problem while Sec. \ref{sec:factor} extends to the
case of multiple assets with identical correlations (e.g., a sector of
the market).  We quantify the improvement in terms of the expected
Sharpe ratio (or risk-adjusted return) of the portfolio and the Sharpe
gain compared to a static allocation scheme.  Conclusion section
summarizes the main results, while technical derivations are reported
in Appendices.

\section{Market model and trading strategy}
\label{sec:general}

We first introduce a mathematical market model for $n$ assets and
describe linear trend following strategies.  We then present the
dynamic portfolio allocation based on a linear combination of strategy
signals.  In this frame, we derive mean and variance of portfolio
returns, formulate the optimal allocation problem, and show its
reduction to a standard static allocation problem for $n^2$ virtual
assets.

\subsection{Market model}

We assume that the return%
\footnote{
Throughout this paper, daily price variations are called ``returns''
for the sake of simplicity.  Rigorously speaking, we consider additive
logarithmic returns resized by realized volatility which is a common
practice on futures markets \cite{Martin12,Thomas12}.  Although asset
returns are known to exhibit various non-Gaussian features (so-called
``stylized facts''
\cite{Bouchaud,Mantegna,Mantegna95,Bouchaud01,Sornette03,Bouchaud04}),
resizing by realized volatility allows one to reduce, to some extent,
the impact of changes in volatility and its correlations
\cite{Bouchaud01b,Valeyre13}, and to get closer to the Gaussian
hypothesis of returns \cite{Andersen00}.}
$r_t^j$ of the $j$-th asset at time $t$ has two contributions: an
instantaneous fluctuation (noise) $\ve_t^j$, and a stochastic trend
which in general is given as a linear combination of random
fluctuations $\xi^j_{t'}$,
\begin{equation}
r_t^j = \ve_t^j + \sum\limits_{t'=1}^{t-1} \A^j_{t,t'} \xi_{t'}^j ,
\end{equation}
where the matrix $\A^j$ describes the stochastic trend of the $j$-th
asset, while $\ve_1^j, \ldots, \ve_t^j$ and $\xi_1^j, \ldots, \xi_t^j$
are two sets of independent Gaussian variables with mean zero and the
following covariance structure:
\begin{equation}
\langle \ve_t^j \ve_{t'}^k\rangle = \delta_{t,t'} \C_\ve^{j,k}, \qquad
\langle \xi_t^j \xi_{t'}^k\rangle = \delta_{t,t'} \C_\xi^{j,k}, \qquad
\langle \ve_t^j \xi_{t'}^k\rangle = 0,
\end{equation}
where $\delta_{t,t'} = 1$ for $t = t'$ and $0$ otherwise.  Here
$\C_\ve$ and $\C_\xi$ are the covariance matrices that describe
inter-asset correlations of noises $\ve_t^j$ and of stochastic trend
components $\xi_t^j$, respectively.  This yields the covariance matrix
of Gaussian asset returns to be
\begin{equation}
\label{eq:C}
\C_{t,t'}^{j,k} \equiv \langle r_t^j r_{t'}^k \rangle = \delta_{t,t'} \C_\ve^{j,k} +  \C_{\xi}^{j,k} (\A^j\A^{k,\T})_{t,t'} ,
\end{equation}
where $\T$ denotes the matrix transposition.  For each asset, the
stochastic trend induces auto-correlations due to a linear combination
of exogenous random variables $\xi_t^j$ which are independent from
short-time noises $\ve_t^j$.  Moreover, the structure of these {\it
auto-correlations} (which is described by the matrix $\A^j$) is
considered to be independent from {\it inter-asset correlations}
(which are described by matrices $\C_\ve$ and $\C_\xi$).  In
particular, the covariance matrices $\C_\ve$ and $\C_\xi$ do not
depend on time.  As discussed in \cite{Grebenkov14}, the presence of
auto-correlations makes TF strategies profitable even for assets with
zero mean returns.  In other words, we consider asset
auto-correlations as the origin of profitability of TF strategies.
Although the whole analysis can be performed for nonzero mean returns,
it is convenient to impose $\langle \ve_t^j\rangle = \langle
\xi_t^j\rangle = \langle r_t^j\rangle = 0$ in order to accentuate the
gain of the TF strategy over a simple buy-and-hold strategy (which is
profitless in this case).

\subsection{Profit-and-loss of a TF portfolio}

The incremental profit-and-loss of a TF portfolio (i.e., the total
return of the portfolio at time $t$) is
\begin{equation}
\dPNL_t = \sum\limits_{j=1}^n r_t^j ~ S^j_{t-1} ,
\end{equation}
where $S^j_{t-1}$ is the position%
\footnote{
The term ``position'' refers to the exposure or investment in a given
asset.  It is generally used in futures trading where position can be
either positive (long) or negative (short) \cite{Hull}.}
of the TF strategy on the $j$-th asset at time $t-1$.  In a
conventional setting, the position $S^j_{t-1}$ is determined from
earlier returns $r^j_1$, ..., $r^j_{t-1}$ of the $j$-th asset.  In
this paper, we will show that this conventional choice is suboptimal
due to inter-asset corrections.  To overcome this limitation, we
introduce the position $S^j_{t-1}$ as a weighted linear combination of
the signals from all assets:
\begin{equation}
S^j_{t-1} = \sum\limits_{k=1}^n \omega_{j,k} ~ s^k(r_1^k,\ldots, r_{t-1}^k) ,
\end{equation}
where $s^k(r_1^k,\ldots,r_{t-1}^k)$ is the signal from the $k$-th
asset, with weights $\omega_{j,k}$ to be determined.  Note that the
weights are considered to be time-independent, in coherence with the
earlier assumption of time-independent inter-asset correlations.  The
incremental P\&L of the portfolio becomes
\begin{equation}
\label{eq:dPNL}
\dPNL_t = \sum\limits_{j,k=1}^n \omega_{j,k} ~r_t^j ~ s^k(r_1^k,\ldots, r_{t-1}^k) ,
\end{equation}
where $\omega_{j,k}$ can be interpreted as the weight of the $k$-th
signal onto the position of $j$-th asset.  The particular case of
diagonal weights (when $\omega_{j,k} = 0$ for $j\ne k$) corresponds to
a portfolio of $n$ TF strategies with weights $\omega_{j,j}$.
Therefore, the standard portfolio allocation problem is included in
our framework, in which the diagonal weight $\omega_{j,j}$ represents
the amount of capital allocated to the $j$-th asset.  In general,
non-diagonal terms allow one to benefit from inter-asset correlations
to enhance the profitability of the TF portfolio.

Following \cite{Grebenkov14}, we consider a TF strategy whose signal
is determined by a {\it linear} combination of earlier returns (e.g.,
an exponential moving average, see below):
\begin{equation}
s^k(r_1^k,\ldots, r_{t-1}^k) = \sum\limits_{t'=1}^{t-1} \S^k_{t,t'} r_{t'}^k ,
\end{equation}
so that
\begin{equation}
\label{eq:dPNL_general}
\dPNL_t = \sum\limits_{j,k=1}^n  \omega_{j,k} \sum\limits_{t'=1}^{t-1} \S^k_{t,t'} r_t^j r_{t'}^k ,
\end{equation}
with given matrices $\S^k_{t,t'}$.

Using the Gaussian character of the model, we compute in
\ref{sec:mean_var} the mean and variance of this incremental
profit-and-loss of a portfolio with $n$ assets:
\begin{equation}
\label{eq:mean_var}
\begin{split}
\langle \dPNL_t \rangle & = \sum\limits_{j,k=1}^n  \omega_{j,k} ~M^{j,k}_t,  \\
\var\{ \dPNL_t \} & = \sum\limits_{j_1,k_1,j_2,k_2=1}^n  \omega_{j_1,k_1} \omega_{j_2,k_2} V^{j_1,k_1;j_2,k_2}_t, \\
\end{split}
\end{equation}
where
\begin{equation}
\label{eq:MV_general}
\begin{split}
M^{j,k}_t & = \C_{\xi}^{j,k} (\S^k \A^k\A^{j,\T})_{t,t} , \\
V^{j_1,k_1; j_2,k_2}_t & = \C_\ve^{j_1,j_2} \C_\ve^{k_1,k_2} (\S^{k_1} \S^{k_2,\T})_{t,t} + 
 \C_\ve^{j_1,j_2} \C_{\xi}^{k_1,k_2} (\S^{k_1} \A^{k_1} \A^{k_2,\T} \S^{k_2,\T})_{t,t} \\
& + \C_\ve^{k_1,k_2}  \C_\xi^{j_1,j_2} (\S^{k_1} \S^{k_2,\T})_{t,t} (\A^{j_1}\A^{j_2,\T})_{t,t} \\
& +  \C_\xi^{j_1,j_2} (\A^{j_1}\A^{j_2,\T})_{t,t} \C_{\xi}^{k_1,k_2} 
(\S^{k_1} \A^{k_1} \A^{k_2,\T} \S^{k_2,\T})_{t,t}   \\
& + \C_{\xi}^{j_1,k_2} \C_{\xi}^{k_1,j_2} (\S^{k_1} \A^{k_1,\T} \A^{j_2})_{t,t}  
(\S^{k_2} \A^{j_1,\T} \A^{k_2})_{t,t} . \\
\end{split}
\end{equation}
The structural separation between auto-correlations and inter-asset
corrections from Eq. (\ref{eq:C}) is also reflected in these formulas.

\subsection{Optimization problem}

Once the mean and variance of the incremental P\&L are derived in the
form (\ref{eq:mean_var}), the dynamic allocation problem for a
portfolio of trend following strategies is reduced to the standard
optimization problem for a portfolio composed of $n^2$ ``virtual''
assets (indexed by a double index $j,k$) whose means are $M^{j,k}_t$
and the covariance is $V^{j_1,k_1;j_2,k_2}_t$.  One can therefore
search for the weights $\omega_{j,k}$ that optimize the chosen
criterion (e.g., to minimize variance under a fixed expected return
for the Markowitz theory).  In this work, we search for the optimal
weights $\omega_{j,k}$ that maximize the squared Sharpe ratio (or
squared risk-adjusted return of the portfolio):
\begin{equation}
\label{eq:Sharpe}
\Sh^2 \equiv \frac{\langle \dPNL_t \rangle^2}{\var\{ \dPNL_t \}} = \frac{(M_t^\T \omega)^2}{(\omega^\T V_t \omega)} ,
\end{equation}
where the weights $\omega_{j,k}$ are denoted here by a single vector
$\omega$ of size $n^2$.  The optimization equations are obtained by
setting
\begin{equation}
\frac{\partial \Sh^2}{\partial \omega_{j,k}} = \frac{2(M_t^\T \omega)}{(\omega^\T V_t \omega)^2} 
\bigl[ M_t^{j,k}  (\omega^\T V_t \omega) - (V_t \omega)^{j,k} (M_t^\T \omega)\bigr] = 0   \qquad (j,k =1,\ldots, n),
\end{equation}
and we used the symmetry of the matrix $V_t$: $V_t^{j_1,k_1;j_2,k_2} =
V_t^{j_2,k_2;j_1,k_1}$.  More explicitly, these equations read
\begin{equation}
\label{eq:optim_equations}
\sum\limits_{j_1,k_1,j_2,k_2=1}^n \bigl[M_t^{j,k} V_t^{j_1,k_1;j_2,k_2} - V_t^{j,k;j_1,k_1} M_t^{j_2,k_2} \bigr]
\omega_{j_1,k_1} \omega_{j_2,k_2} = 0 
\end{equation}
for all indices $j,k=1,\ldots,n$.  In general, this is a set of $n^2$
quadratic equations onto $n^2$ unknown weights $\omega_{j,k}$.
However, the original expressions for the mean and variance of
$\dPNL_t$ are invariant under the substitution of $\omega_{j,k}$ by
$\omega_{k,j}$.  This is related to the linearity of the considered
trend following strategy.  In what follows, we consider the symmetric
weights so that there remain $n(n+1)/2$ unknown weights, with the same
number of equations.  For instance, one needs to solve $3$, $6$ and
$10$ equations for a portfolio with two, three and four assets,
respectively.  Note also that any solution of the above optimization
problem is defined up to a multiplicative factor.  In fact, the
squared Sharpe ratio in Eq. (\ref{eq:Sharpe}) is invariant under
multiplication of weights $\omega_{j,k}$ by any nonzero constant.  As
a consequence, $\omega_{j,k}$ should be interpreted as relative
weights.

In general, the solution of Eqs. (\ref{eq:optim_equations}) depends on
two covariance matrices $\C_\ve$ and $\C_\xi$ (inter-asset
correlations), matrices $\A^j$ (asset auto-correlations), and matrices
$\S^j$ (signals of TF strategies).  Once all these matrices are
specified, the optimization problem can be solved numerically.
However, it is impossible in practice to infer such a large number of
parameters from market data, as well as to understand their influences
on the optimal weights.  For this reason, we further specify the
problem in order to reduce the original, very large set of parameters.
First, we choose in Sec. \ref{sec:EMA} a particular form of
auto-correlations (matrices $\A^j$) and TF signals (matrices $\S^j$).
After that, we consider several particular forms of the covariances
matrices $\C_\ve$ and $\C_\xi$ for two-asset case and a sector model.
In this way, we identify a small number of the most relevant
parameters and investigate their influence onto the optimal TF
portfolio.

\subsection{Exponential moving averages}
\label{sec:EMA}

In \cite{Grebenkov14}, we employed exponential moving averages (EMAs)
to describe both stochastic trends and signals of TF strategies.  This
is equivalent to choosing stochastic trends as induced by a discrete
Ornstein-Uhlenbeck process for which
\begin{equation}
\A^j_{t,t'} = \begin{cases}  \beta^j (1-\lambda^j)^{t-t'-1} , \quad t > t' , \cr 0 , \hskip 30mm t \leq t',  \end{cases}
\end{equation}
where $\beta^j$ and $\lambda^j$ are the strength and the rate of the
$j$-th stochastic trend.  Similarly, the signal of a TF strategy is
also chosen to be an EMA \cite{Winters60,Brown}:
\begin{equation}
\S^j_{t,t'} = \begin{cases}  \gamma^j (1-\eta^j)^{t-t'-1} , \quad t > t' , \cr 0 , \hskip 29mm t \leq t',  \end{cases}
\end{equation}
where $\gamma^j$ and $\eta^j$ are the strength and the rate of the
$j$-th TF strategy.  Setting the elements of these matrices to $0$ for
$t \leq t'$ implements the causality: the trend and the signal at time
$t$ rely only upon the earlier returns with $t' < t$.

In what follows, we focus on the particular situation when the rates
$\lambda^j$ of all assets are identical ($\lambda^j = \lambda$), and
the rates $\eta^j$ of all strategies are identical ($\eta^j = \eta$).
In the stationary limit $t\to\infty$, we derive in
\ref{sec:mean_var}: 
\begin{equation}
\label{eq:M_V_st2}
\begin{split}
M_\infty^{j,k} & = \frac{q \sqrt{1-p^2}}{(1-pq)(1-q^2)}~ \C_{\xi,\beta}^{j,k} , \\
V_\infty^{j_1,k_1; j_2,k_2} & = \C_\ve^{j_1,j_2} \C_\ve^{k_1,k_2} + \frac{2 \C_\ve^{j_1,j_2} \C_{\xi,\beta}^{k_1,k_2}}{(1-pq)(1-q^2)} 
 + \C_{\xi,\beta}^{j_1,j_2} \C_{\xi,\beta}^{k_1,k_2} \frac{1+q^2 - 2p^2q^2}{(1-pq)^2(1-q^2)^2} ,  \\
\end{split}
\end{equation}
where $q = 1-\lambda$, $p = 1-\eta$, and we set $\gamma^k = \gamma =
\sqrt{1-p^2}$ as an appropriate normalization (see \cite{Grebenkov14}).

In general, the optimal weights maximizing the squared Sharpe ratio
can be found by solving numerically either the set of
Eqs. (\ref{eq:optim_equations}), or the unconstrained maximization
problem for $\Sh^2$ in Eq. (\ref{eq:Sharpe}).  In order to understand
the mechanisms behind the optimal TF strategy, we first focus on the
particular case of two assets, for which many results can be derived
analytically and then easily illustrated (Sec. \ref{sec:two}).  After
that, we consider in Sec. \ref{sec:factor} a sector model of $n$
similar assets.

\section{Two assets}
\label{sec:two}

For two assets, there are three independent weights: $\omega_{11}$,
$\omega_{22}$, and $\omega_{12}$.  The covariance matrices take a
simple form:
\begin{equation}
\C_\ve = \left(\begin{array}{c c} \sigma^1 \sigma^1 & \sigma^1 \sigma^2 \rho_\ve \\ \sigma^1 \sigma^2 \rho_\ve & \sigma^2 \sigma^2 \end{array}\right) , \qquad
\C_\xi = \left(\begin{array}{c c} 1 & \rho_\xi \\ \rho_\xi & 1 \end{array}\right) ,
\end{equation}
where $\rho_\ve$ and $\rho_\xi$ are two correlation coefficients
(between inter-asset price noises $\ve^1_t$, $\ve^2_t$, and stochastic
trend components $\xi^1_t$, $\xi^2_t$, respectively), while $\sigma^1$
and $\sigma^2$ are the volatilities of noises $\ve^j_t$.  Note that
the volatility of stochastic trend components $\xi^j_t$ can be
included into auto-correlation strengths $\beta^j$ that allows one to
write a simplified form of the covariance matrix $\C_\xi$.

Substituting these relations in Eq. (\ref{eq:M_V_st2}), we get
explicit formulas for the mean and variance of the incremental
profit-and-loss $\dPNL_\infty$ in the stationary regime:
\begin{equation}
\label{eq:mean_var_two}
\begin{split}
\langle \dPNL_\infty \rangle & = \frac{q \sqrt{1-p^2}~ [\beta^2_0]^2}{1-pq} \bigl[\kappa^2 \omega_{11} + 2\rho_\xi \kappa \omega_{12} + \omega_{22} \bigr] , \\
\var\{\dPNL_\infty \} & = [\sigma^2]^4 \bigl(\Omega_1 +  2\Omega_2/Q +  R \Omega_3/Q^2 \bigr)  \\
\end{split}
\end{equation} 
(with $\beta_0^{1,2} = \beta^{1,2}/\sqrt{1-q^2}$) so that the squared
Sharpe ratio becomes
\begin{equation}
\label{eq:Sharpe_two}
\Sh^2 = \frac{q^2 (1-p^2) (\kappa^2 \omega_{11} + 2\rho_\xi \kappa \omega_{12} + \omega_{22})^2}{Q^2 \Omega_1 + 2Q\Omega_2 + R\Omega_3} ,
\end{equation}
where
\begin{equation}
\label{eq:Omega}
\begin{split}
\Omega_1 & \equiv \nu^4 \omega_{11}^2 + 4\rho_\ve \nu \omega_{12}[\nu^2 \omega_{11} + \omega_{22}] + 2\nu^2 \rho_\ve^2 \omega_{11}\omega_{22} 
 + 2\nu^2(1+\rho_\ve^2) \omega_{12}^2 + \omega_{22}^2 ,  \\
\Omega_2 & \equiv \nu^2\kappa^2 \omega_{11}^2 + 2\omega_{11}\omega_{12} \nu\kappa (\nu \rho_\xi + \kappa \rho_\ve) 
+ 2\omega_{11}\omega_{22} \nu\kappa \rho_\ve \rho_\xi \\
& + \omega_{12}^2(\nu^2 + 2\nu\kappa \rho_\ve \rho_\xi + \kappa^2) 
+ 2\omega_{12}\omega_{22}(\nu \rho_\ve + \kappa\rho_\xi) + \omega_{22}^2,  \\
\Omega_3 & \equiv \kappa^4 \omega_{11}^2 + 4\rho_\xi \kappa \omega_{12}[\kappa^2 \omega_{11} + \omega_{22}] + 2\kappa^2 \rho_\xi^2 \omega_{11}\omega_{22} 
 + 2\kappa^2(1+\rho_\xi^2) \omega_{12}^2 + \omega_{22}^2 ,  \\
\end{split}
\end{equation} 
and
\begin{equation}
\label{eq:QR}
Q \equiv \frac{(1-pq)[\sigma^2]^2}{[\beta^2_0]^2}, \quad R \equiv 1+q^2-2p^2q^2, \quad  \kappa \equiv \frac{\beta^1}{\beta^2} , 
\quad \nu \equiv \frac{\sigma^1}{\sigma^2} .
\end{equation}
With no loss of generality, we assume that $\beta_1 \leq \beta_2$,
i.e. $\kappa \leq 1$.

As discussed in Sec. \ref{sec:general}, the optimization procedure to
maximize $\Sh^2$ leads to three quadratic equations on weights
$\omega_{11}$, $\omega_{12}$, and $\omega_{22}$.  Defining two
independent ratios,
\begin{equation}
z = \frac{\omega_{11}}{\omega_{22}} , \qquad x = \frac{\omega_{12}}{\omega_{22}}, 
\end{equation}
one gets three equations containing terms $z^2$, $x^2$, $zx$, $z$,
$x$, and constants.  The equation $\frac{\partial
\Sh^2}{\partial \omega_{11}} = 0$ does not contain the term $z^2$, while
the equation $\frac{\partial \Sh^2}{\partial \omega_{12}} = 0$ does
not contain the term $x^2$.  Taking appropriate linear combinations,
one can express and then eliminate the term $xz$.  Finally, one would
deal with a single fourth degree equation.  Although an explicit
analytical solution of this equation is possible, it is too cumbersome
to any practical use.  In turn, the original problem of maximizing the
squared Sharpe ratio can be solved numerically as a standard
minimization problem.

The squared Sharpe ratio in Eq. (\ref{eq:Sharpe_two}) depends on the
following parameters of the model: two rates $\lambda (=1-q)$ and
$\eta (=1-p)$ of the EMAs for stochastic trends and for TF strategies;
two asset volatilities $\sigma^1$ and $\sigma^2$; two auto-correlation
strengths $\beta^1$ and $\beta^2$; and two correlation coefficients
$\rho_\ve$ and $\rho_\xi$.  In order to illustrate and discuss various
features of the optimal solution, we consider several particular cases
of practical interest for which explicit analytical solutions are
relatively simple.

\subsection{Uncorrelated assets ($\rho_\ve = \rho_\xi = 0$)}
\label{sec:uncorrelated}

We first consider the case of two uncorrelated assets: $\rho_\ve =
\rho_\xi = 0$.  In this case, the condition $\frac{\partial
\Sh^2}{\partial \omega_{12}} = 0$ leads to the equation $(\nu^2Q^2 +
Q(\nu^2+\kappa^2) +\kappa^2R) xz = 0$.  Since the coefficient in front
of $xz$ is strictly positive, one has either $x=0$, or $z=0$.  The
second option ($z=0$) does not satisfy other equations while the
former case yields
\begin{equation}
\label{eq:xz_opt_uncorr}
x_{\rm opt} = \frac{\omega_{12}}{\omega_{22}} = 0, \qquad 
z_{\rm opt} = \frac{\omega_{11}}{\omega_{22}} = \frac{\kappa^2(Q^2 + 2Q + R)}{\nu^4 Q^2 + 2\nu^2 \kappa^2 Q + \kappa^4 R},
\end{equation}
with the squared optimal Sharpe ratio
\begin{equation}
\label{eq:Sopt_uncorr}
\Sh^2_{\rm opt} = q^2(1-p^2) \biggl[\frac{1}{Q^2 + 2Q + R} + \frac{\kappa^4}{\nu^4 Q^2 + 2\nu^2 \kappa^2 Q + \kappa^4 R} \biggr] .
\end{equation}

Since two assets are uncorrelated, no additional information can be
gained by including lead-lag term so that $\omega_{12} = 0$, in agreement
with Eq. (\ref{eq:xz_opt_uncorr}).  In turn, $z_{\rm opt}$ determines
the weights of two assets in the optimal portfolio, up to a
multiplicative factor.  Adding a constrain $\omega_{11} + \omega_{22} = 1$, one
can identify $\omega_{11}$ and $\omega_{22}$ are relative weights.

When two assets have identical characteristics (i.e., $\sigma^1 =
\sigma^2$ and $\beta^1 = \beta^2$ from which $\nu = \kappa = 1$),
Eq. (\ref{eq:xz_opt_uncorr}) yields $z_{\rm opt} = 1$, i.e., both
assets enter with equal weights as expected.  In this case, the
squared Sharpe ratio of the optimal portfolio is twice larger than the
Sharpe ratio of either asset: $\Sh^2_{\rm opt}(\kappa=1) = 2
\frac{q^2(1-p^2)}{Q^2+2Q+R}$, in agreement with the
principle of diversification.

\begin{figure}
\begin{center}
\includegraphics[width=0.49\textwidth]{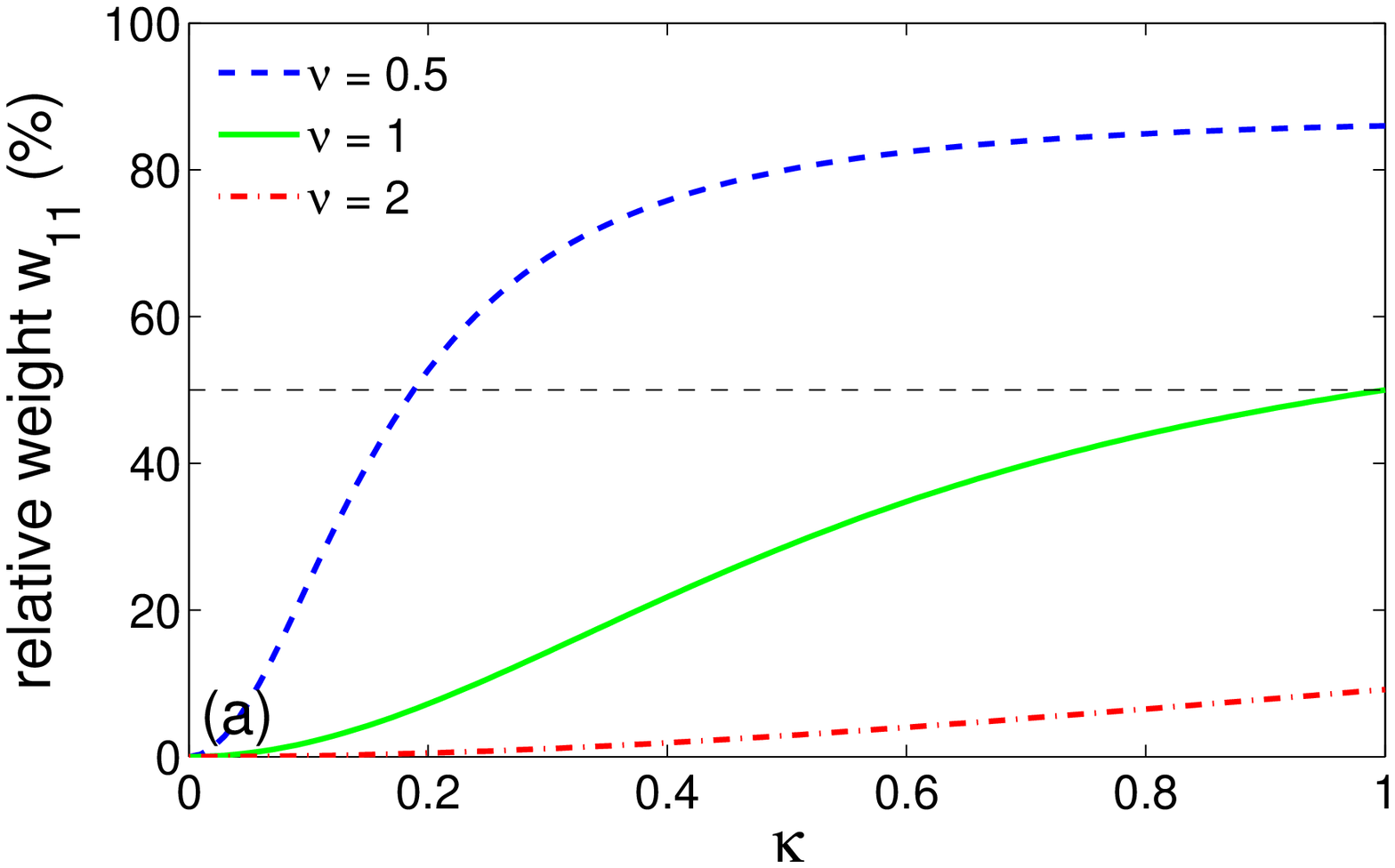}  
\includegraphics[width=0.49\textwidth]{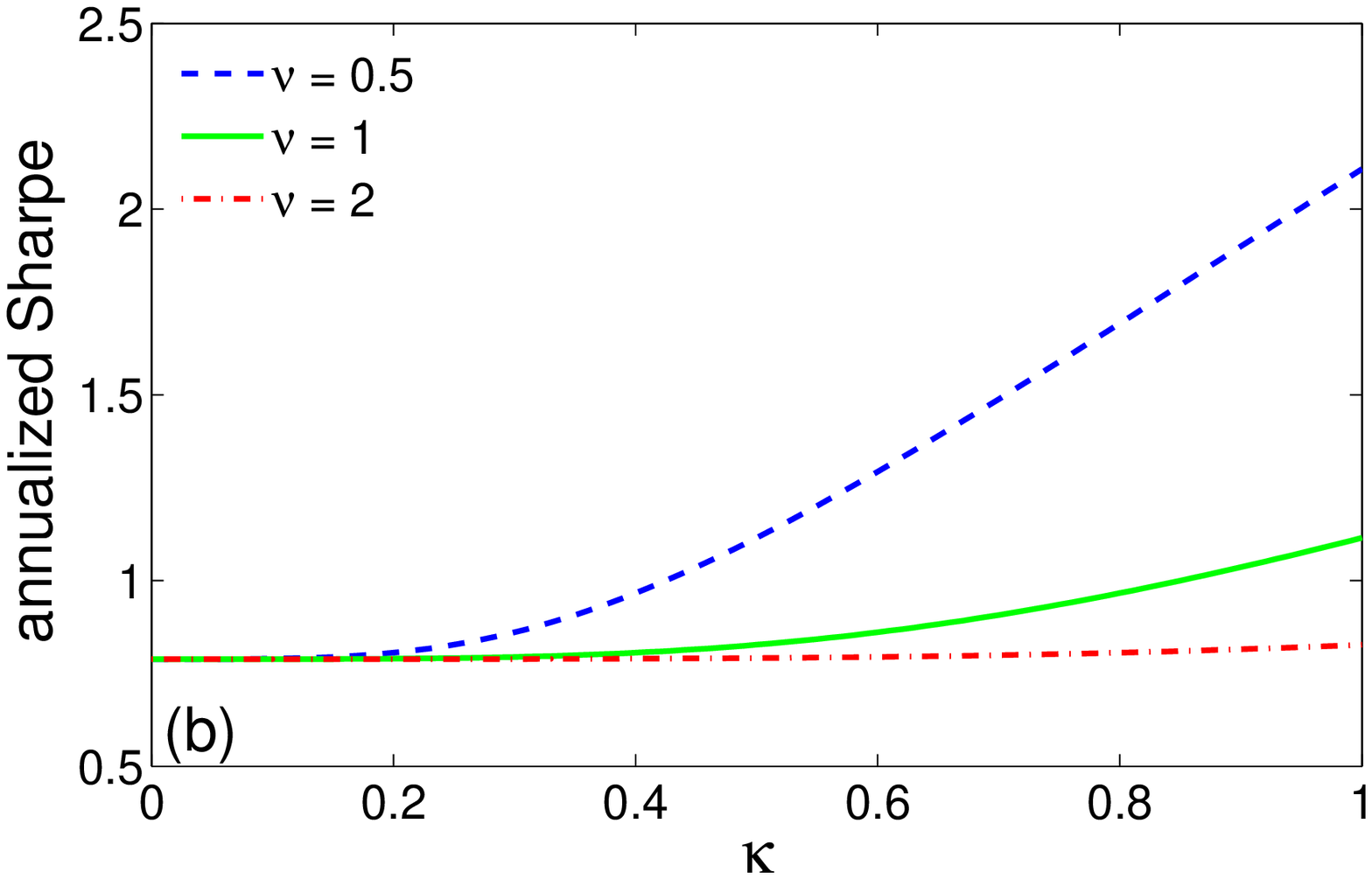}  
\end{center}
\caption{
The relative weight $\omega_{11} = 100\% ~\frac{z_{\rm opt}}{1 +
z_{\rm opt}}$ of the first asset in the optimal portfolio ({\bf a})
and the annualized optimal Sharpe ratio $\sqrt{255}~ \Sh_{\rm opt}$
({\bf b}), versus $\kappa = \beta^1_0/\beta^2_0$, for uncorrelated
assets (i.e., $\rho_\ve = \rho_\xi = 0$), with $\beta^2_0 = 0.1$,
$\beta^1_0 = \kappa \beta^2_0$, $\lambda = \eta = 0.01$, $\sigma^2 =
1$, and $\sigma^1 = \nu \sigma^2$, with $\nu = 0.5, 1, 2$.  }
\label{fig:uncorr}
\end{figure}

In the opposite limit $\kappa = 0$, the first asset has no
auto-correlation ($\beta^1 = 0$) so that the underlying TF strategy is
profitless and therefore excluded from the portfolio: $\omega_{11} = z_{\rm
opt} = 0$.  One retrieves the squared Sharpe ratio of the second
asset: $\Sh^2_{\rm opt}(\kappa = 0) = \frac{q^2(1-p^2)}{Q^2+2Q+R}$.  
Figure \ref{fig:uncorr} shows the relative weight $\omega_{11} = 100\%
\frac{z_{\rm opt}}{1+z_{\rm opt}}$ of the first asset in the optimal
portfolio and the annualized%
\footnote{
Since we investigate the incremental profit-and-loss $\dPNL_t$, the
Sharpe ratio $\Sh$ in Eq. (\ref{eq:Sharpe}) characterizes the
risk-adjusted return of the portfolio over one time step of TF
strategies.  In many trading platforms, TF strategies are realized on
a daily basis (i.e., they are updated once per day).  At the same
time, it is conventional to rescale the {\it daily} Sharpe ratio $\Sh$
to the {\it annualized} Sharpe ratio $\sqrt{255}~\Sh$, where
$\sqrt{255}$ is the standard pre-factor matching a calendar year of
255 business days.  For a quick grasp of this rescaling, one can use
as a reference point an asset delivering a 10\% annual return with
10\% annualized volatility, corresponding to a Sharpe ratio of $1$.
In the systematic hedge industry, only a few funds deliver the Sharpe
ratio above 1 over the long run \cite{Thomas12,Burghardt10,Lo02}.}
optimal Sharpe ratio as functions of $\kappa$ varying from $0$ to $1$.
For two assets with the same volatility (i.e., $\nu = 1$), the
relative weight $\omega_{11}$ of the less profitable first asset
varies from $0$ to $50\%$, while the squared Sharpe ratio doubles, as
expected.  When the first (less profitable) asset is in addition twice
more volatile ($\nu = 2$), its relative weight does not exceed $10\%$
even at $\kappa = 1$, while the Sharpe ratio has almost not improved
(dash-dotted lines).  In turn, if the less profitable asset is twice
less volatile ($\nu = 0.5$), its relative weight rapidly grows and
exceeds $50\%$ at $\kappa \simeq 0.18$.  The inclusion of the less
volatile asset greatly improves the Sharpe ratio.  Using
Eqs. (\ref{eq:xz_opt_uncorr}, \ref{eq:Sopt_uncorr}), one can
investigate the relative impacts of volatilities and auto-correlations
onto the optimal portfolio for uncorrelated assets.

\subsection{Indistinguishable correlated assets ($\kappa = \nu = 1$)}
\label{sec:Stwo_indist}

In order to reveal the role of inter-asset corrections, we consider
two assets with the same structure of auto-correlations (i.e., $\kappa
= \nu = 1$) that makes them indistinguishable from each other.  Since
each asset offers the same expected TF returns, the diagonal weights
are expected to be identical: $\omega_{11} = \omega_{22}$.  We study
therefore the optimal value of the lead-lag correction $\omega_{12}$
(which is the same for both assets), and the gain in the Sharpe ratio
that this cross-correcting term brings to the optimal portfolio.

In \ref{sec:two_indist}, we derive the optimal solution of this
minimization problem: $z_{\rm opt} = 1$,
\begin{equation}
\label{eq:xopt_indist}
x_{\rm opt} =  - \frac{Q^2(2\rho_\ve - \rho_\xi - \rho_\ve^2 \rho_\xi) + 2Q\rho_\ve(1-\rho_\xi^2) + R\rho_\xi(1-\rho_\xi^2)}
{Q^2(1-2\rho_\ve \rho_\xi + \rho_\ve^2) + 2Q(1-\rho_\xi^2) + R(1-\rho_\xi^2)} ,
\end{equation}
and the squared optimal Sharpe ratio is
\begin{equation}
\label{eq:Sopt_indist}
\begin{split}
\Sh^2_{\rm opt} & = 2q^2(1-p^2) \biggl[(1-\rho_\xi^2)\biggl(Q^2(1-\rho_\ve^2)+2Q(1-\rho_\ve \rho_\xi)+R(1-\rho_\xi^2)\biggr) \\
& +2 Q^2(\rho_\ve-\rho_\xi)^2\biggr] \biggl\{\biggl(Q^2(1-\rho_\ve^2)+4Q(1-\rho_\ve\rho_\xi)+R(1-\rho_\xi^2)\biggr) \\
& \times \biggl(Q^2(1-\rho_\ve^2)+R(1-\rho_\xi^2)\biggr) +4Q^2(1-\rho_\ve^2)(1-\rho_\xi^2)+4Q^2R(\rho_\ve-\rho_\xi)^2 \biggr\}^{-1} . \\
\end{split}
\end{equation}
In the typical situation, $R\ll Q$ (see Eq. (\ref{eq:QR})), and if
$\rho_\ve$ is not too close to $1$, one can neglect terms with $R$ in
order to get a simpler approximate relation:
\begin{equation}
\begin{split}
\Sh^2_{\rm opt} & \simeq 2q^2(1-p^2) \frac{1-\rho_\xi^2}{1-\rho_\ve^2} ~
\frac{Q[1-\rho_\ve^2 + 2(\rho_\ve-\rho_\xi)^2] + 2(1-\rho_\ve\rho_\xi)}{Q^2(1-\rho_\ve^2) + 4Q(1-\rho_\ve\rho_\xi)+4(1-\rho_\xi^2)} . \\
\end{split}
\end{equation}

In order to assess the gain of including the lead-lag term, one can
compare $\Sh_{\rm opt}$ with the Sharpe ratio $\Sh_0$ from
Eq. (\ref{eq:Sharpe_two}) with $\omega_{11} = \omega_{22}$ and
$\omega_{12} = 0$ (i.e., without lead-lag term):
\begin{equation}
\label{eq:S0}
\Sh^2_0 = \frac{q^2(1-p^2)}{Q^2(1+\rho_\ve^2) + 2Q(1+\rho_\ve\rho_\xi) + R(1+\rho_\xi^2)} .
\end{equation}

\begin{figure}
\begin{center}
\includegraphics[width=0.49\textwidth]{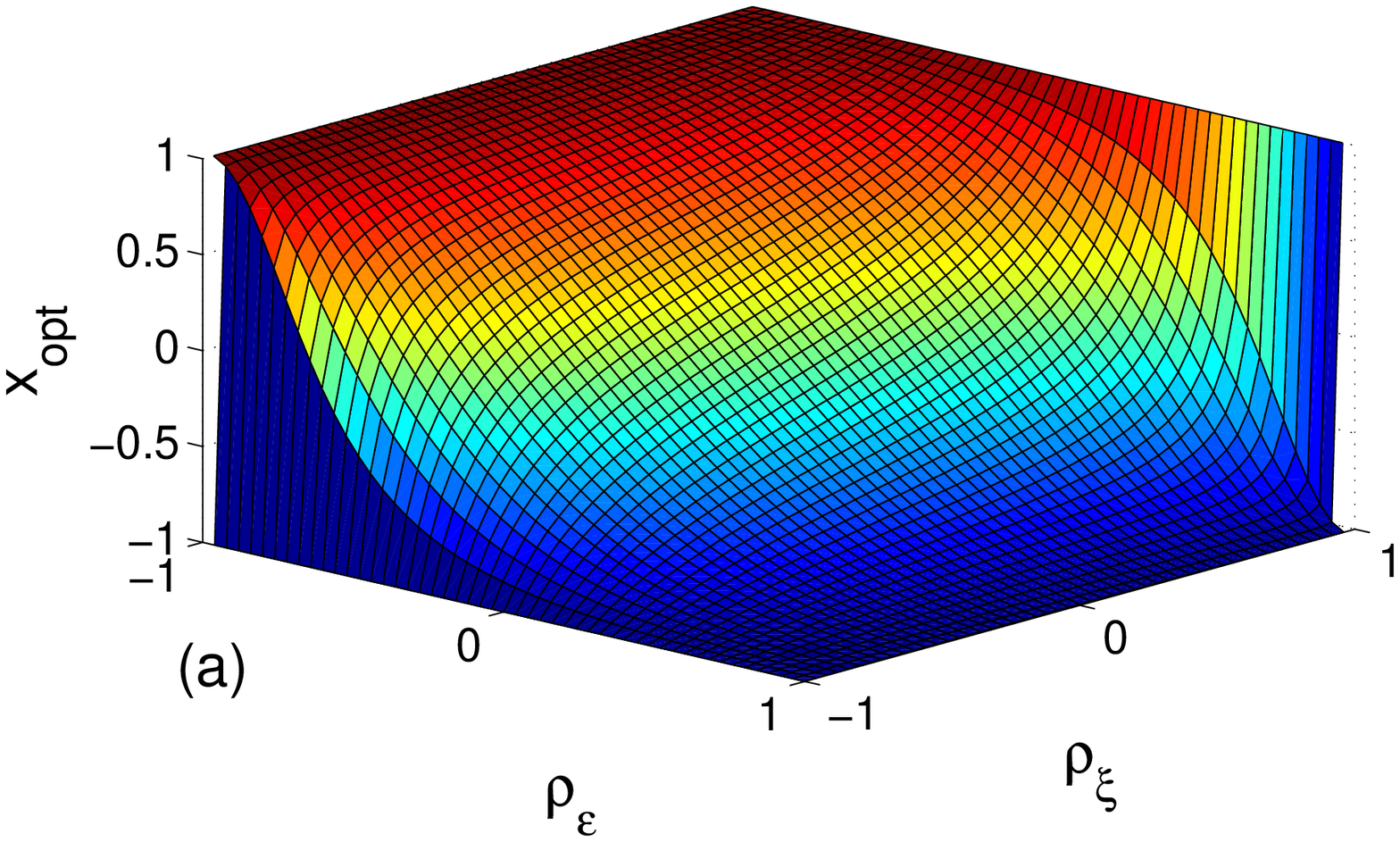}  
\includegraphics[width=0.49\textwidth]{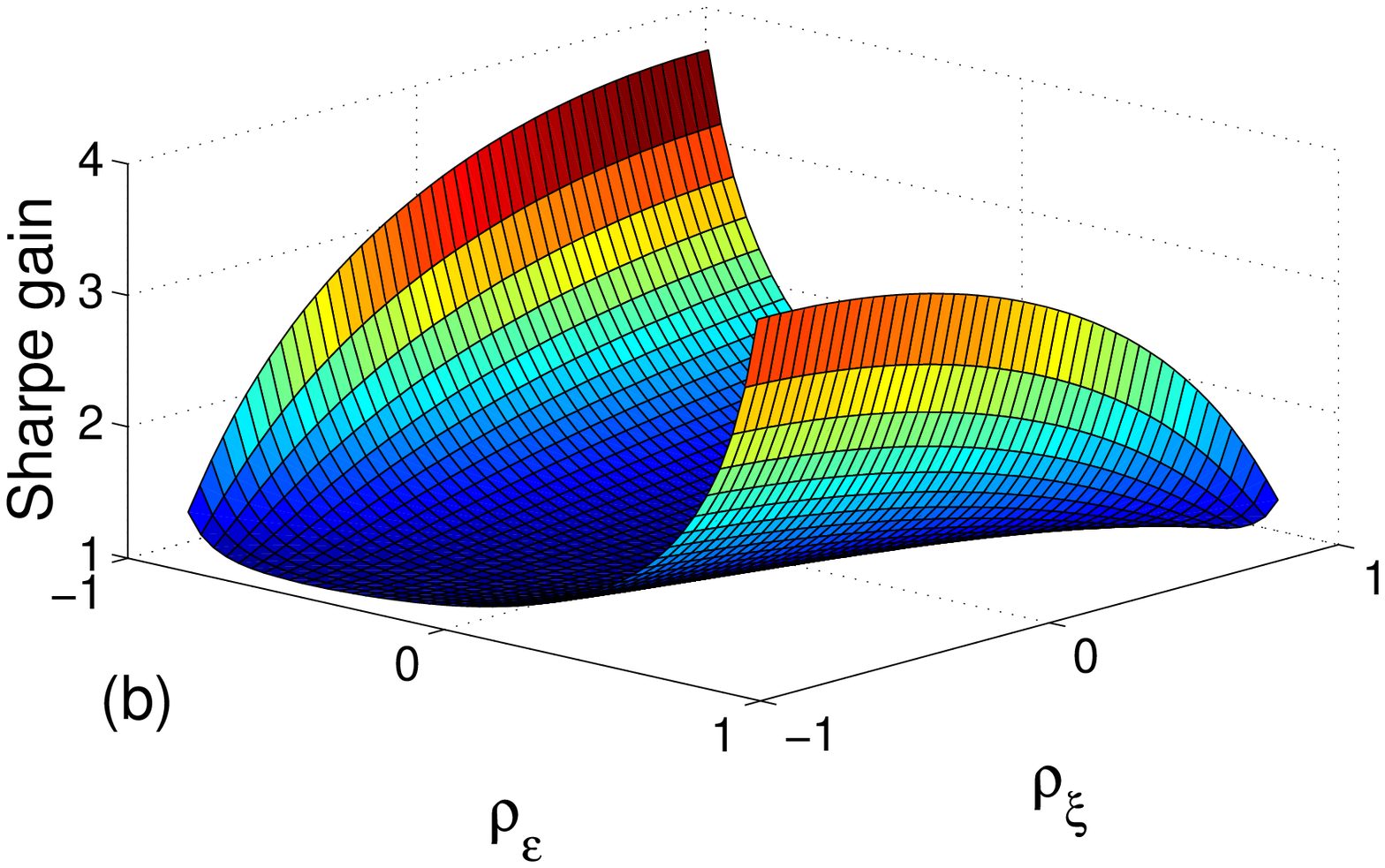}  
\end{center}
\caption{
The optimal lead-lag weight ratio $x_{\rm opt} =
\omega_{12}/\omega_{22}$ {\bf (a)} and the Sharpe gain $\Sh_{\rm
opt}/\Sh_0$ {\bf (b)} as functions of two correlation coefficients
$\rho_\ve$ and $\rho_\xi$ for two indistinguishable assets with
$\beta^1_0 = \beta^2_0 = 0.1$, $\sigma^1 = \sigma^2 = 1$, and $\lambda
= \eta = 0.01$.  For the second plot, both correlation coefficients
vary from $-0.9$ to $0.9$ in order to exclude unrealistically large
Sharpe gains at $\rho_\ve$ and $\rho_\xi$ around $\pm 1$.}
\label{fig:Sopt_3D}
\end{figure}

Figure \ref{fig:Sopt_3D} shows how $x_{\rm opt}$ and the Sharpe gain
$\Sh_{\rm opt}/\Sh_0$ of the optimal portfolio depend on two
correlation coefficients $\rho_\ve$ and $\rho_\xi$.  As expected, the
highest gain can be achieved when $\rho_\ve \rho_\xi$ is close to
$-1$, e.g., when stochastic trends are strongly correlated ($\rho_\xi
\simeq 1$) while noises are strongly anti-correlated ($\rho_\ve \simeq
-1$).  This extreme situation illustrates the need to distinguish
correlations in trends and noises.  In traditional portfolio
optimization which only operates with inter-asset correlations, the
explicit separation of the two effects is not possible.  As a
consequence, significant increases in Sharpe ratio can be overlooked
in such models.  At the same time, a reliable estimation of two
correlation coefficients from financial data remains challenging.

In order to better grasp the behavior of $x_{\rm opt}$ and $\Sh_{\rm
opt}$, we consider several particular cases:

(i) when $\rho_\xi = \pm 1$ (fully correlated stochastic trends), one
gets $x_{\rm opt} = \pm 1$, independently of $\rho_\ve$.  As seen on
Fig. \ref{fig:Sopt_3D}a, this behavior is unstable: it is sufficient
to take $|\rho_\xi|$ slightly smaller than $1$ to retrieve the
dependence of $x_{\rm opt}$ on $\rho_\ve$.

(ii) when $\rho_\ve = \pm 1$ (fully correlated noises), one gets
$x_{\rm opt} \simeq \pm 1$, with a very weak dependence on $\rho_\xi$.
In contrast to the above case, this behavior persists for all
$\rho_\ve$ near $\pm 1$ (Fig. \ref{fig:Sopt_3D}a).

(iii) when $\rho_\ve = 0$ (uncorrelated noises), one finds
\begin{equation}
\label{eq:xopt_option1}
x_{\rm opt} = \rho_\xi \frac{Q^2 - R(1-\rho_\xi^2)}{Q^2 + (2Q+R)(1-\rho_\xi^2)} \approx \frac{\rho_\xi}{1 + (1-\rho_\xi^2)\frac{2}{Q}}  ,
\end{equation}
where the small term $R$ was neglected in the second relation.  Here
one can see the impact of correlated trends on apparently uncorrelated
markets.  Note that the lead-lag correction has the same sign as the
trend correlation coefficient $\rho_\xi$.

(iv) when $\rho_\xi = 0$ (uncorrelated stochastic trends), one gets
\begin{equation}
\label{eq:xopt_option2}
x_{\rm opt} = - \rho_\ve \frac{2Q(Q+1)}{Q^2(1+\rho_\ve^2) + 2Q+R} \approx \frac{-\rho_\ve}{1 - (1-\rho_\ve^2)\frac{Q}{2(Q+1)}} ,
\end{equation}
where the small term with $R$ was neglected in the second relation.
One can see the impact of correlated noises on a market with
independent trend components.  In contrast to the above case, the
lead-lag correction has the opposite sign of the noise correlation
coefficient $\rho_\ve$.

(v) when $\rho_\ve = \rho_\xi = \rho$, one gets $x_{\rm opt} = -\rho$,
i.e., the position of the first asset should be reduced by a relative
amount of $\rho$ of the second asset in order to maximally decorrelate
them.  In addition, the squared optimal Sharpe ratio does not depend
on correlations:
\begin{equation}
\Sh^2_{\rm opt} = \frac{2q^2(1-p^2)}{Q^2+2Q+R} .
\end{equation}
In other words, such correlations cannot improve the optimal Sharpe
ratio but one needs to correct the TF strategy to remove the effect of
correlations.  This case is particularly interesting as it helps to
show that the static allocation is suboptimal without introducing the
lead-lag correction.

\begin{figure}
\begin{center}
\includegraphics[width=0.49\textwidth]{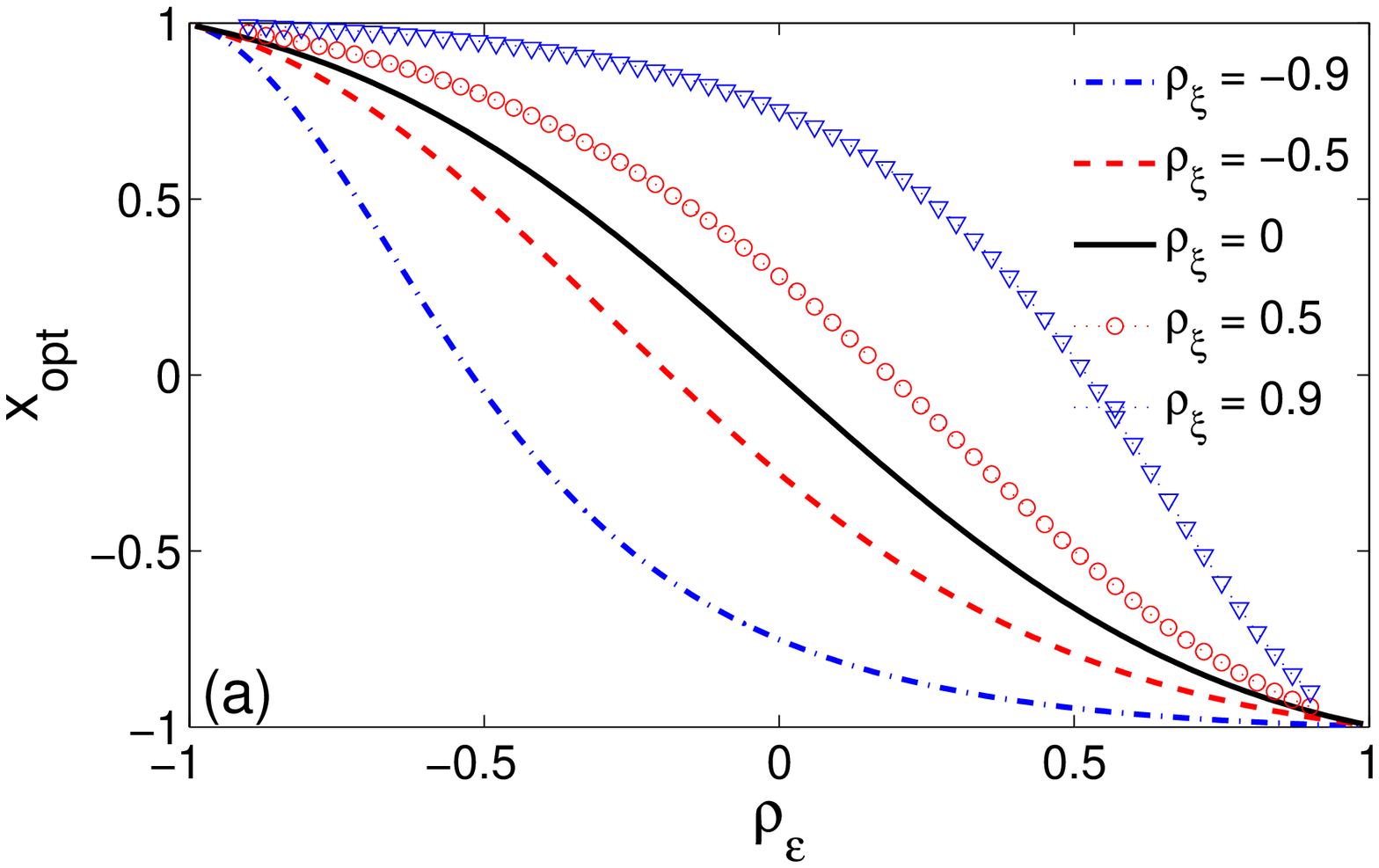} 
  \includegraphics[width=0.49\textwidth]{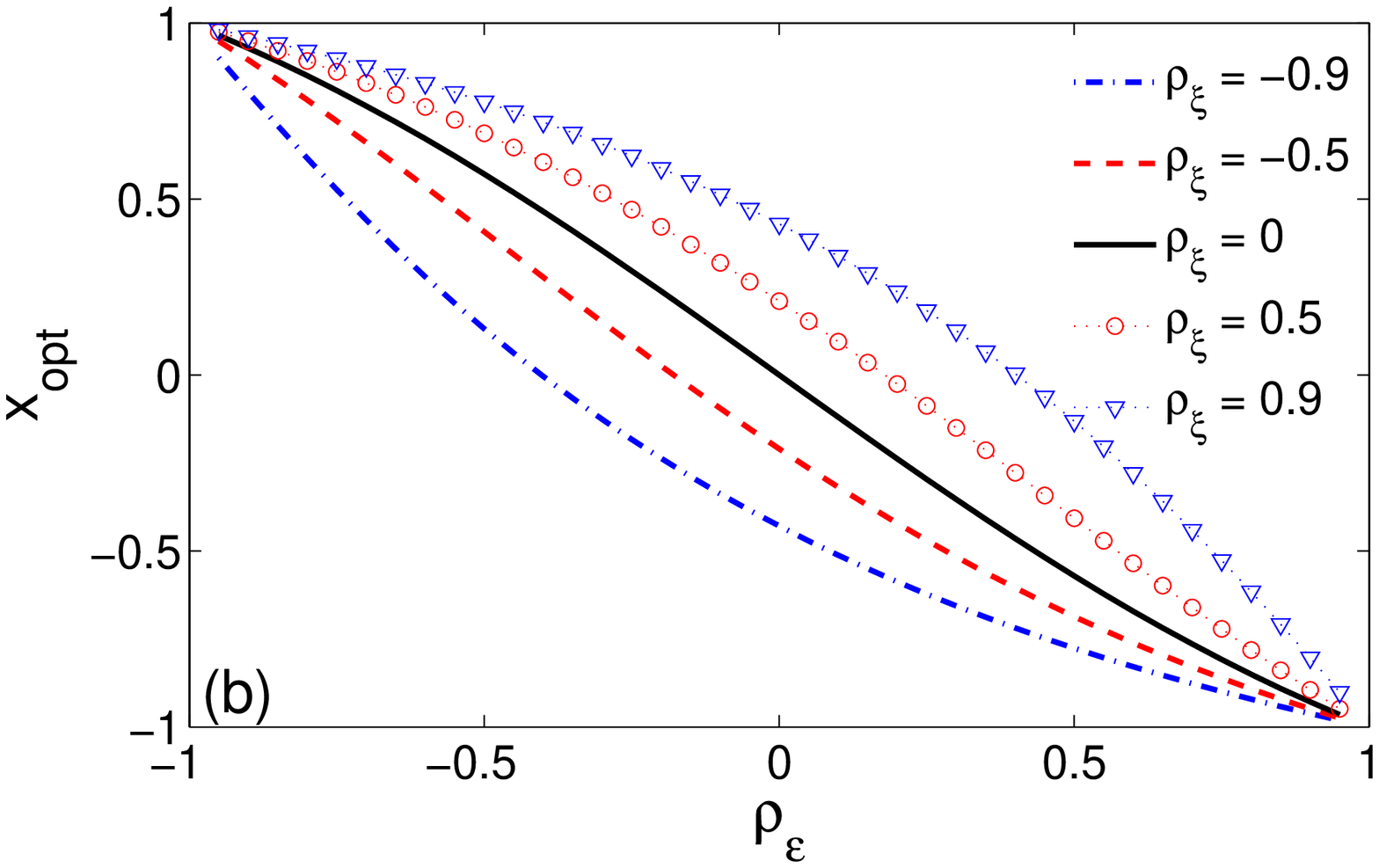}  
\includegraphics[width=0.49\textwidth]{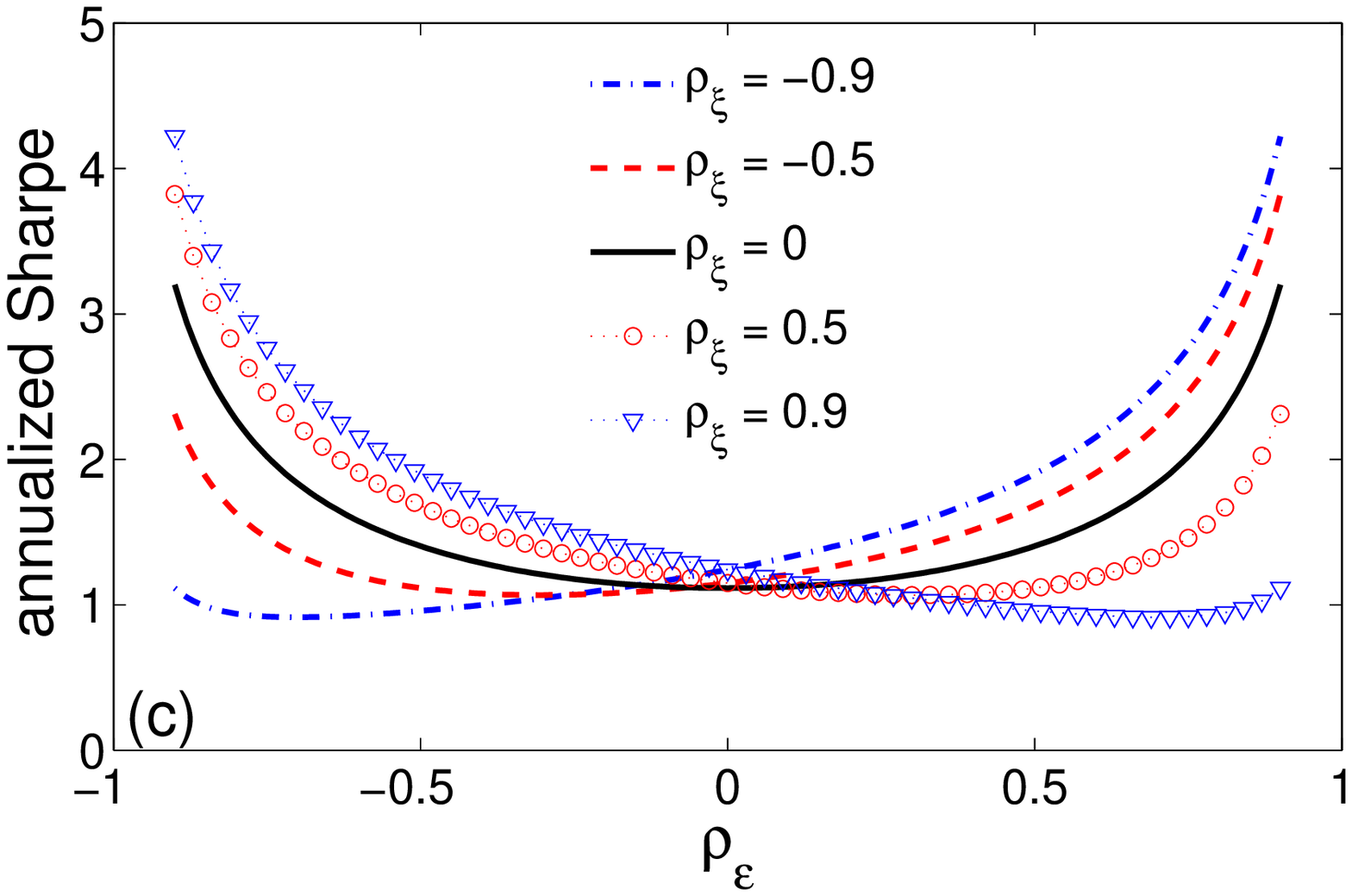}    
  \includegraphics[width=0.49\textwidth]{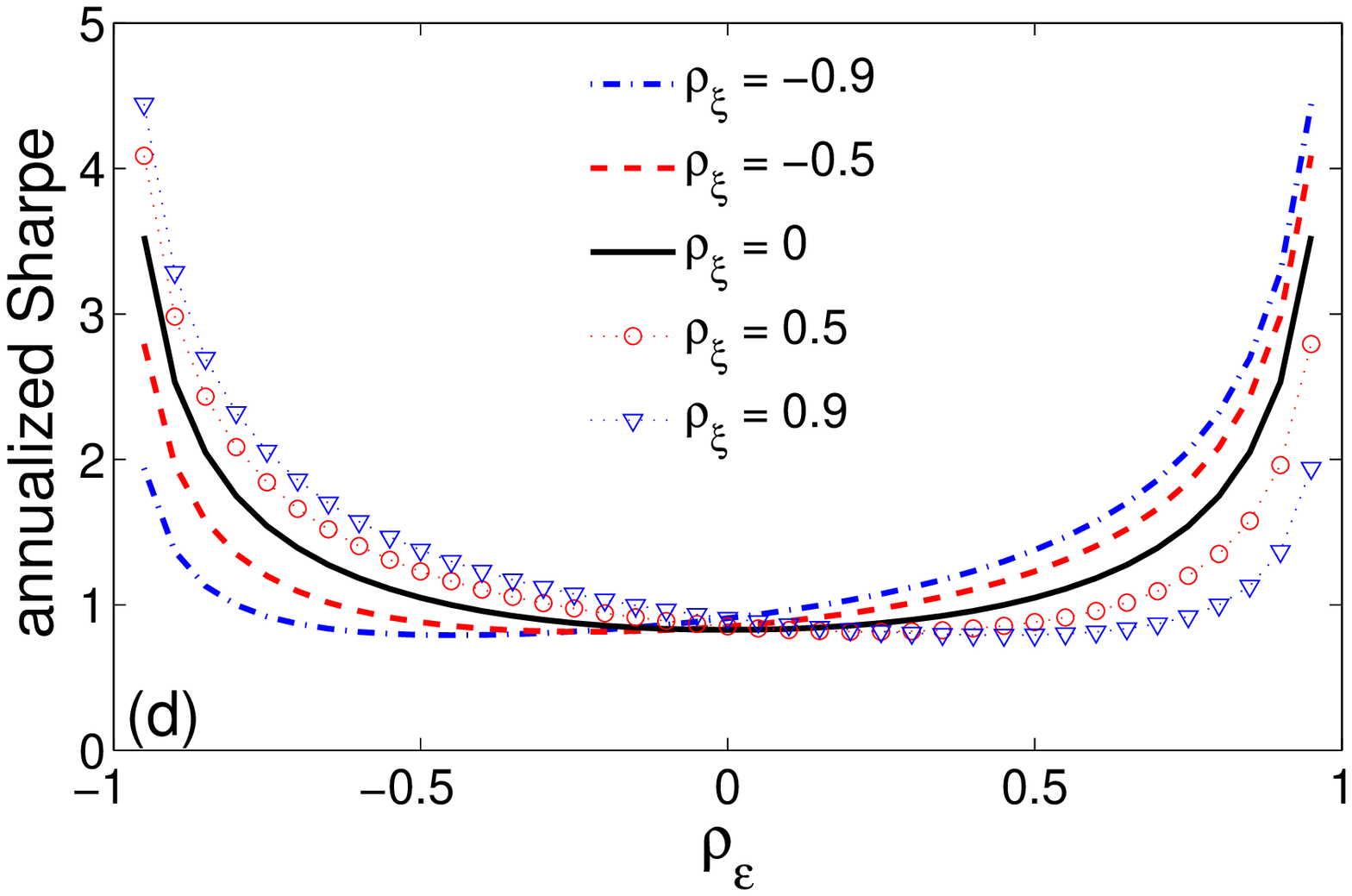}  
\includegraphics[width=0.49\textwidth]{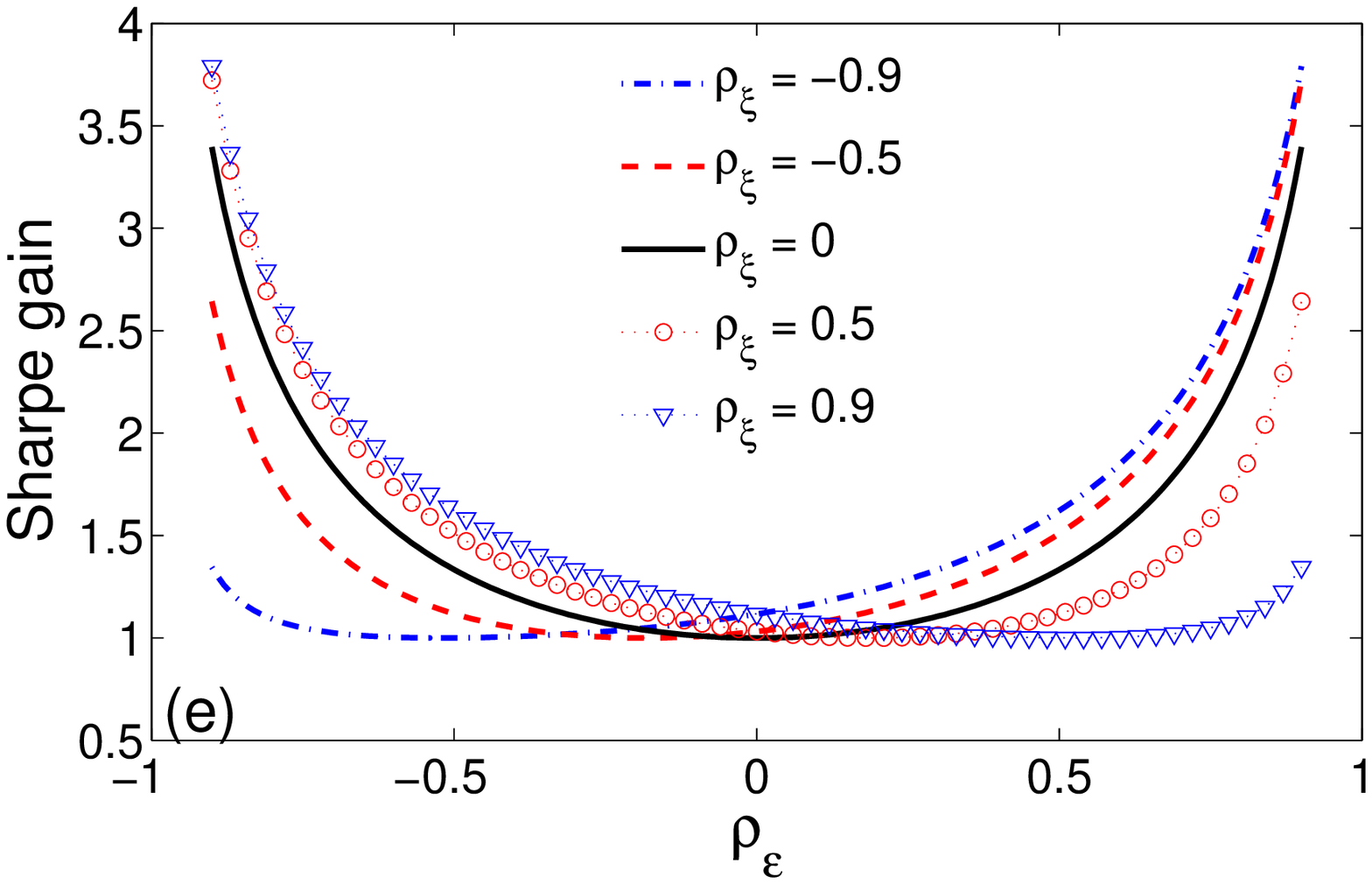}   
  \includegraphics[width=0.49\textwidth]{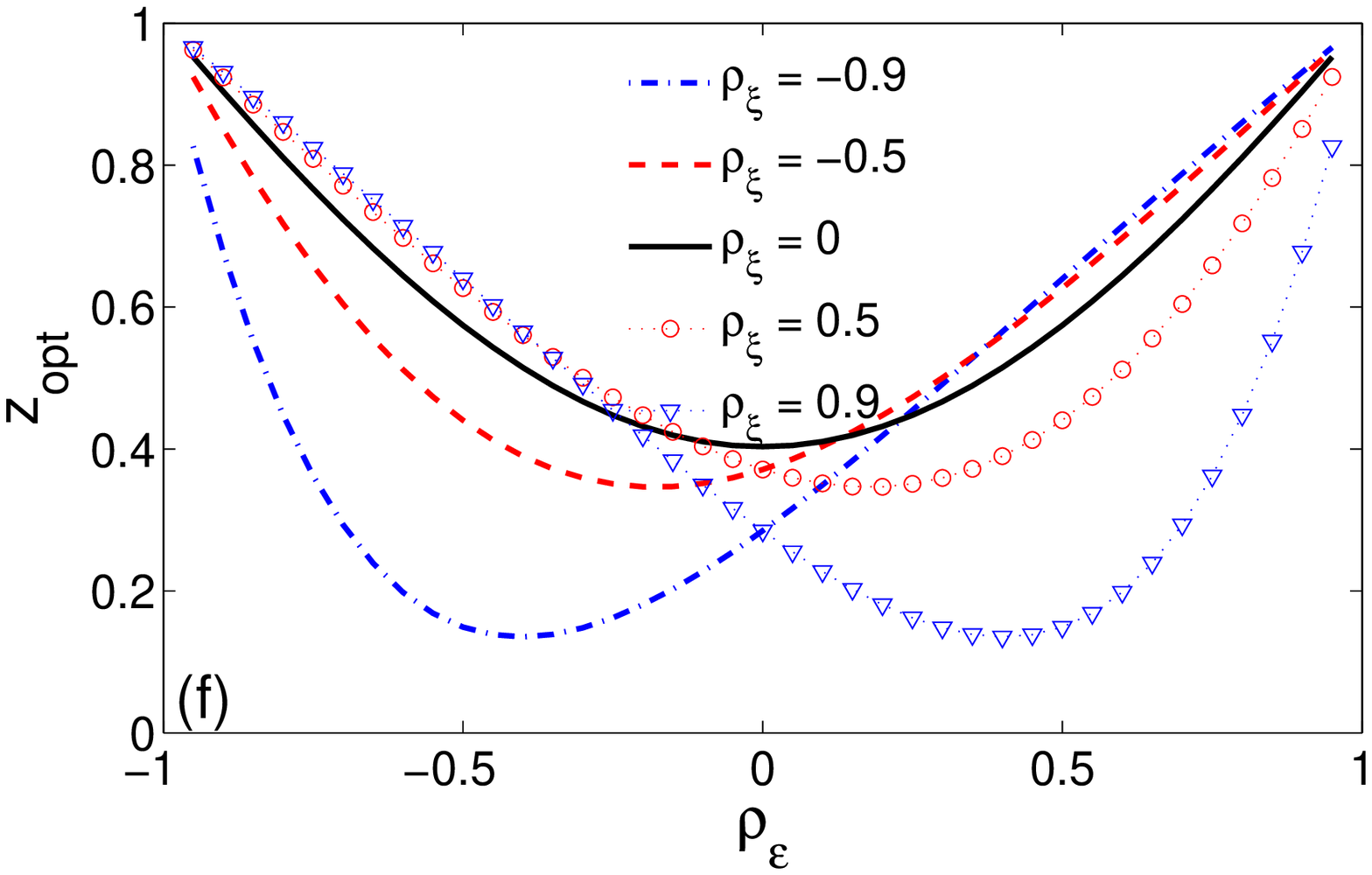}  
\end{center}
\caption{
Comparison between optimal portfolios for two indistinguishable assets
($\beta^1_0 = \beta^2_0 = 0.1$, $\kappa = 1$, left) and two assets
with different stochastic trends ($\beta^1_0 = 0.05$, $\beta^2_0 =
0.1$, $\kappa = 0.5$, right): the optimal lead-lag correction $x_{\rm
opt} = \omega_{12}/\omega_{22}$ ({\bf a,b}), the annualized optimal
Sharpe ratio $\sqrt{255}~ \Sh_{\rm opt}$ ({\bf c,d}), the gain in the
Sharpe ratio $\Sh_{\rm opt}/\Sh_0$ due to the lead-lag correction
({\bf e}), and the optimal asset weights ratio $z_{\rm opt} =
\omega_{11}/\omega_{22}$ ({\bf f}) (note that $z_{\rm opt} = 1$ for
indistinguishable assets on the left).  These quantities are presented
as functions of $\rho_\ve$ for different $\rho_\xi$, and we set
$\sigma^1 = \sigma^2 = 1$ (i.e., $\nu = 1$) and $\lambda = \eta =
0.01$.  }
\label{fig:Sopt}
\end{figure}

The plots in the left column of Fig. \ref{fig:Sopt} further illustrate
some properties of the optimal portfolio of TF strategies on two
indistinguishable assets for different correlation coefficients
$\rho_\ve$ and $\rho_\xi$.  Both assets are traded with identical
weights, $\omega_{11} = \omega_{22}$, i.e. $z_{\rm opt} = 1$.  The
optimal lead-lag correction $x_{\rm opt} = \omega_{12}/\omega_{22}$
from Eq. (\ref{eq:xopt_indist}) monotonously decreases with $\rho_\ve$
from $1$ at $\rho_\ve = -1$ (fully anticorrelated noises) to $-1$ at
$\rho_\ve = 1$ (fully correlated noises), as shown on
Fig. \ref{fig:Sopt}a.  The rate of decrease depends on $\rho_\xi$.  As
expected, no correction is needed when $\rho_\ve = \rho_\xi = 0$.
Generally, the horizontal line at $x_{\rm opt} = 0$ determines the set
of correlation coefficients for which no lead-lag correction term is
needed.

Figure \ref{fig:Sopt}c shows how the annualized optimal Sharpe ratio
$\sqrt{255} ~\Sh_{\rm opt}$ changes with correlation coefficients
$\rho_\ve$ and $\rho_\xi$.  When $\rho_\ve = 0$, the annualized Sharpe
ratio is close $1$ (for the chosen level $\beta_0 = 0.1$ of
auto-correlations), and it depends weakly on stochastic trend
correlations ($\rho_\xi$).  In turn, this ratio is strongly enhanced
when $|\rho_\ve| > 0.5$, and the enhancement occurs when $\rho_\ve$
and $\rho_\xi$ are of opposite signs.  In contrast, the annualized
optimal Sharpe ratio may be decreased when both correlations are of
the same sign.  Finally, the enhancement is even stronger when
$|\rho_\ve|$ is close to $1$, independently of mutual signs of
$\rho_\ve$ and $\rho_\xi$.  However, this region seems to be
unrealistic for markets.

Figure \ref{fig:Sopt}e illustrates the Sharpe gain $\Sh_{\rm
opt}/\Sh_0$ due to inclusion of the lead-lag correction term.  This
gain is substantial for highly correlated assets (when $|\rho_\ve|$
and/or $|\rho_\xi|$ are large).  As expected, the gain is always
greater than (or equal to) $1$, as $\Sh_{\rm opt}$ is the optimal
solution.

\subsection{Example of two distinct assets: $\kappa = 0.5$}

For comparison, we present on the right column of Fig. \ref{fig:Sopt}
similar quantities for two assets with different stochastic trends
($\beta^1_0 = 0.05$, $\beta^2_0 = 0.1$, i.e., $\kappa = 0.5$).  In
this example, the first asset with lower auto-correlations is less
profitable for TF strategies.  Although the inclusion of the first
asset does not improve the mean profit-and-loss, it still allows to
increase the Sharpe ratio by reducing the variance due to
diversification.  Figure \ref{fig:Sopt}f shows the optimal weights
ratio $z_{\rm opt} = \omega_{11}/\omega_{22}$.  When there is no
correlations ($\rho_\ve = \rho_\xi = 0$), Eq. (\ref{eq:xz_opt_uncorr})
yields $z_{\rm opt} \approx 0.40$, i.e., two assets enter with
relative weights $28.6\%$ and $71.4\%$, respectively (for chosen
parameters).  For correlated assets, the optimal weights ratio can be
either smaller or larger than $0.4$.  For instance, when both $\rho_\ve$
and $\rho_\xi$ are of the same sign, the relative weight of the first
asset can be reduced to almost zero ($z_{\rm opt}$ is close to $0$).
In contrast, when $|\rho_\ve|$ is close to $1$, both assets have to be
included with almost the same weights ($z_{\rm opt}$ approaches $1$).

Figures \ref{fig:Sopt}b,d show the optimal lead-lag correction $x_{\rm
opt}$ and the annualized optimal Sharpe ratio.  Both quantities
exhibit similar features as in the case of indistinguishable assets
(left column).  The Sharpe gain is also similar to the earlier case
(not shown).

\section{A sector model}
\label{sec:factor}

The above analysis can be extended to multiple assets.  Although the
theoretical solution is formally available, a large number of weights
(growing as $n(n+1)/2$) makes challenging its investigation in
general.  At the same time, one can still perform numerical
minimization to determine the optimal solution by standard
optimization tools.  In this section, we consider the particular case
of a market sector when inter-asset correlation is the same for all
assets \cite{Sharpe}.  We also assume that all returns are normalized
by realized volatilities, i.e., $\sigma^j = 1$.  In other words, we
consider the covariance matrices for noises and stochastic trends to
be
\begin{equation}
\C_\ve = \left(\begin{array}{c c c c c} 
    1    & \rho_\ve & \rho_\ve & ... & \rho_\ve \\
\rho_\ve &     1    & \rho_\ve & ... & \rho_\ve \\
\rho_\ve & \rho_\ve &     1    & ... & \rho_\ve \\
  ...    &     ...  &    ...   & ... &   ...    \\
\rho_\ve & \rho_\ve & \rho_\ve & ... &    1  \\  \end{array}\right), \qquad
\C_\xi = \left(\begin{array}{c c c c c} 
    1    & \rho_\xi & \rho_\xi & ... & \rho_\xi \\
\rho_\xi &     1    & \rho_\xi & ... & \rho_\xi \\
\rho_\xi & \rho_\xi &     1    & ... & \rho_\xi \\
  ...    &     ...  &    ...   & ... &   ...    \\
\rho_\xi & \rho_\xi & \rho_\xi & ... &    1  \\  \end{array}\right) .
\end{equation}
Each of these matrices has two eigenvalues, $\nu_1 = 1 + (n-1)\rho$
and $\nu_2 = 1-\rho$.  Since covariance matrices must be positive
definite, all eigenvalues should be non-negative that implies
$\rho_\ve \geq - 1/(n-1)$ and $\rho_\xi \geq -1/(n-1)$.  In what
follows, we only consider $\rho_\ve \geq 0$ and $\rho_\xi \geq 0$.

When the assets are indistinguishable (i.e., $\beta^j = \beta$), they
are expected to have the same weights in the optimal portfolio of TF
strategies, $\omega_{jj} = \omega_{11}$, as well as all lead-lag
corrections are the same: $\omega_{jk} = \omega_{12}$ for all $j\ne
k$.  For this particular case of indistinguishable assets, the number
of unknown weights is reduced from $n(n-1)/2$ to $2$ that allows one
to derive an analytical solution and to investigate its behavior as a
function of the number of assets.

In \ref{sec:Afactor}, we derived the optimal solution:
\begin{eqnarray}
\label{eq:xopt_multi}
x_{\rm opt} &=& \frac{\omega_{12}}{\omega_{11}} = - \frac{V_2 - (n-1)\rho_\xi V_1}{V_3 - (n-1)\rho_\xi V_2} , \\
\label{eq:Sopt_multi}
\Sh^2_{\rm opt} &=& n q^2 (1-p^2) \frac{(n-1)^2 \rho_\xi^2 V_1 - 2\rho_\xi(n-1) V_2 + V_3}{V_1V_3 - V_2^2} .
\end{eqnarray}
where
\begin{equation}
\label{eq:Vj}
\begin{split}
V_1 & = Q^2(1 + (n-1)\rho_\ve^2) + 2Q(1 + (n-1)\rho_\ve \rho_\xi) + R(1 + (n-1)\rho_\xi^2) , \\ 
V_2 & = (n-1)\bigl[Q^2(2\rho_\ve + (n-2)\rho_\ve^2) + 2Q (\rho_\ve + \rho_\xi + (n-2)\rho_\ve \rho_\xi) \\ 
& + R (2\rho_\xi + (n-2)\rho_\xi^2)\bigr] ,\\ 
V_3 & = n \bigl[Q^2 (1+(n-1)\rho_\ve)^2 + 2Q (1+(n-1)\rho_\ve)(1+(n-1)\rho_\xi) \\ & + R
(1+(n-1)\rho_\xi)^2\bigr] - V_1 - 2V_2. \\
\end{split}
\end{equation}
Although these formulas are explicit, they are rather cumbersome for
theoretical analysis.  For $n = 2$, one retrieves the results of
Sec. \ref{sec:Stwo_indist} for two indistinguishable assets.  In what
follows, we consider several particular cases in order to illustrate
the main features of the optimal solution and the role of the
portfolio size (number of assets).

\subsection{Conventional trading}

It is instructive to start with a ``conventional'' trading without
lead-lag correction term ($\omega_{12} = 0$) for which
Eq. (\ref{eq:Smulti}) yields
\begin{equation}
\Sh^2_n = \frac{n~ q^2(1-p^2)}{Q^2(1 + (n-1)\rho_\ve^2) + 2Q(1 + (n-1)\rho_\ve \rho_\xi) + R(1 + (n-1)\rho_\xi^2)} .
\end{equation}
For $n=1$, one retrieves the squared Sharpe ratio of a single asset:
\begin{equation}
\Sh^2_1 = \frac{q^2(1-p^2)}{Q^2 + 2Q + R}.  
\end{equation}

If there was no inter-asset correlation ($\rho_\ve = \rho_\xi = 0$),
the squared Sharpe ratio $\Sh^2_n$ for $n$ assets would be $n$ times
larger than $\Sh^2_1$ for a single asset: $\Sh^2_n = n \Sh^2_1$, as
expected due to diversification.  In this uncorrelated case, one also
finds the optimal solution to be
\begin{equation}
\label{eq:factor_S0}
x_{\rm opt} = 0, \qquad   \Sh^2_{\rm opt} = \Sh^2_n(\rho_\ve=\rho_\xi=0) = n \Sh^2_1 .
\end{equation}
The presence of correlations reduces the effect of diversification and
diminishes $\Sh_n$.  Moreover, this reduction is stronger for large
$n$.  In what follows, we show that inclusion of the lead-lag term
allows one to recover or even further enhance the Sharpe ratio.  In
other words, although diversification may be reduced by strong
correlations, their proper accounting makes them even more profitable.

\subsection{Equal trend and noise correlations ($\rho_\ve = \rho_\xi$)}

For $\rho_\ve = \rho_\xi = \rho$, we get
\begin{equation*}
\begin{split}
V_1 & = (1+(n-1)\rho^2) [Q^2+2Q+R] , \\
V_2 & = (n-1)(2\rho + (n-2)\rho^2)[Q^2+2Q+R] , \\
V_3 & = (n-1)(1 + 2(n-2)\rho + \rho^2(n^2-3n+3)) [Q^2+2Q+R] . \\
\end{split}
\end{equation*}
Substituting these expressions into Eq. (\ref{eq:Sopt_multi}), we
deduce
\begin{equation}
\label{eq:xopt_Sopt_multi0}
x_{\rm opt} = - \frac{\rho}{1 + (n-2)\rho},  \qquad   \Sh^2_{\rm opt} = n \frac{q^2(1-p^2)}{Q^2+2Q+R} = n \Sh^2_1 .
\end{equation}
As for the case of two assets, the optimal Sharpe ratio does not
depend on correlations, while the lead-lag term does depend on $\rho$.
As previously, one can compare the squared optimal Sharpe ratio to
$\Sh_n^2$ (i.e., the case without lead-lag correction):
\begin{equation}
\frac{\Sh^2_{\rm opt}}{\Sh^2_n} = 1 + (n-1)\rho^2 .
\end{equation}
One can see that accounting for correlations by inclusion of the
lead-lag term may significantly increase the squared Sharpe ratio.
This effect obviously disappears at $n = 1$.  Note that even weak
correlations can be enhanced by including many assets in a portfolio.
This secondary effect (in addition to diversification that increases
$\Sh_n$) favors large portfolios, in agreement with a common practice
of fund managers.

\subsection{Uncorrelated stochastic trends ($\rho_\xi = 0$)}
\label{sec:factor_noise}

\begin{figure}
\begin{center}
\includegraphics[width=0.49\textwidth]{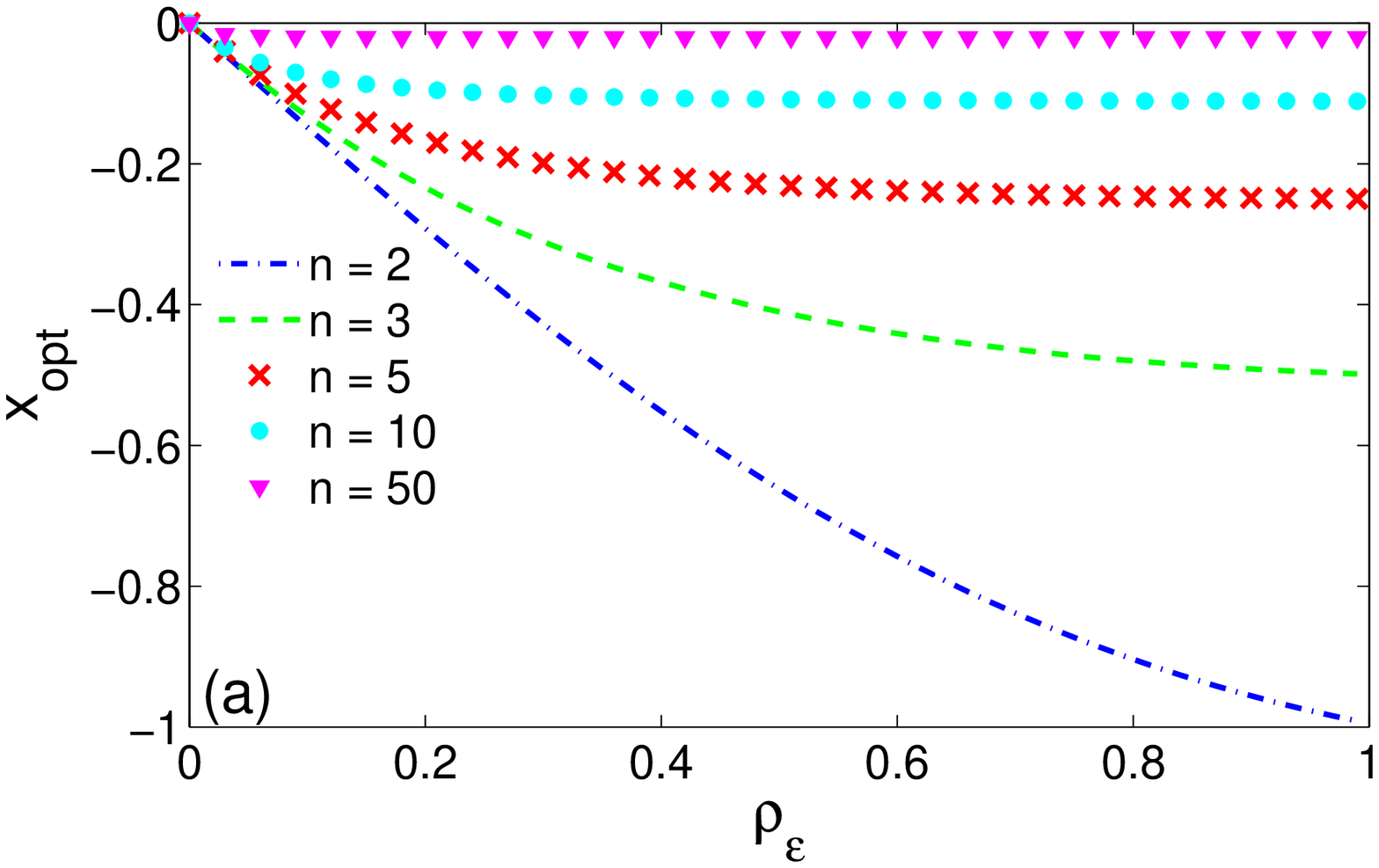}  
\includegraphics[width=0.49\textwidth]{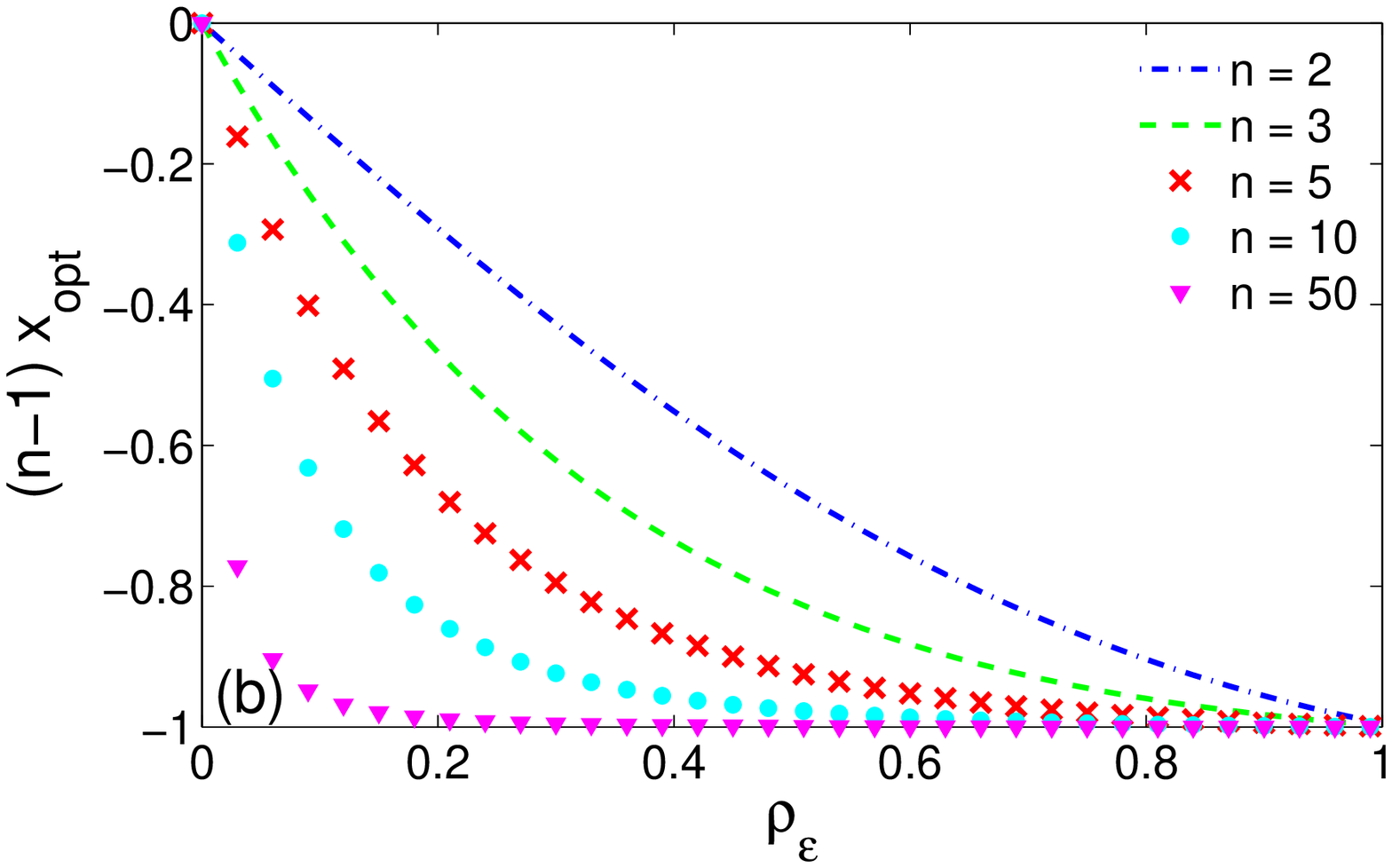}  
\includegraphics[width=0.49\textwidth]{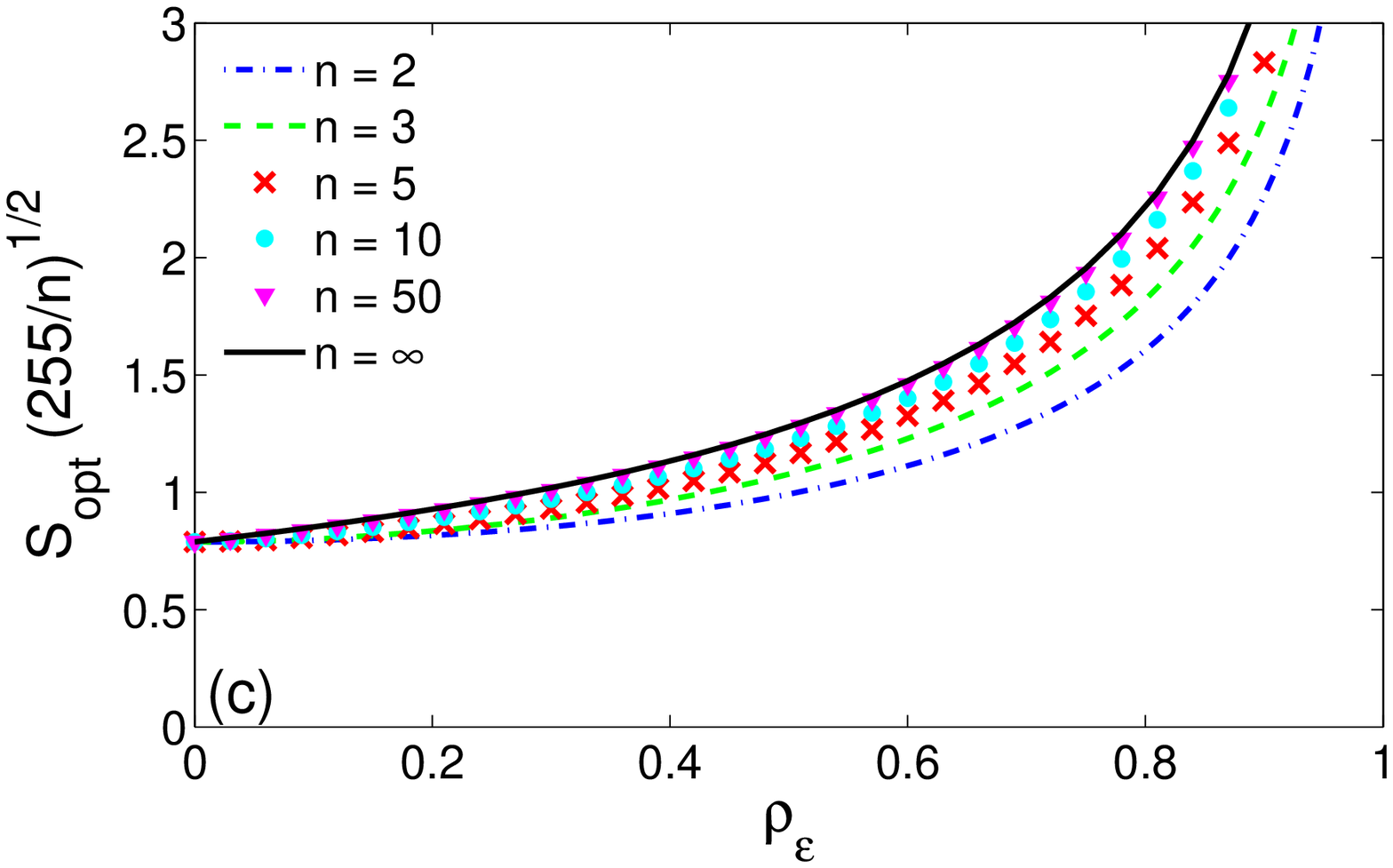}  
\includegraphics[width=0.49\textwidth]{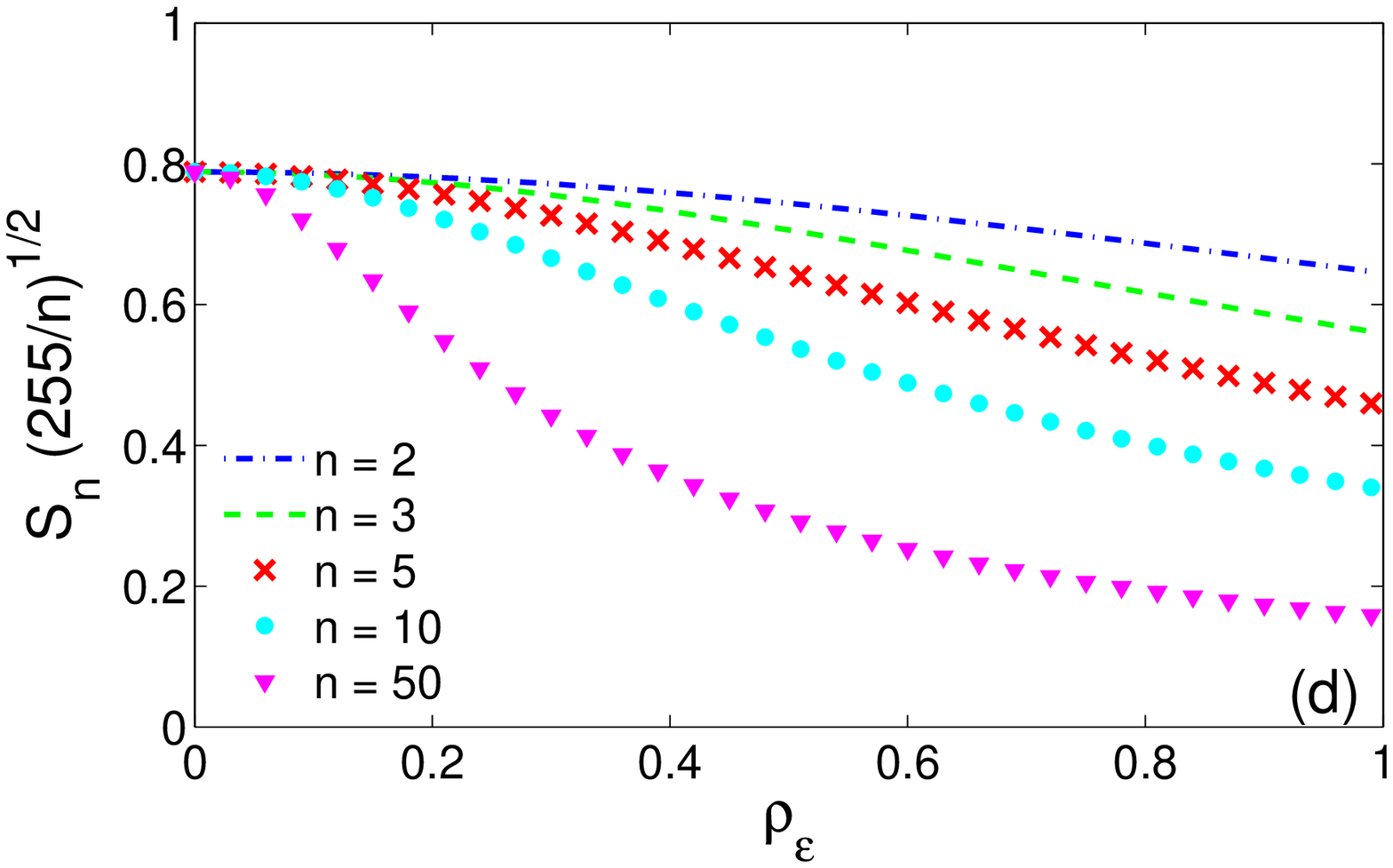}  
\includegraphics[width=0.49\textwidth]{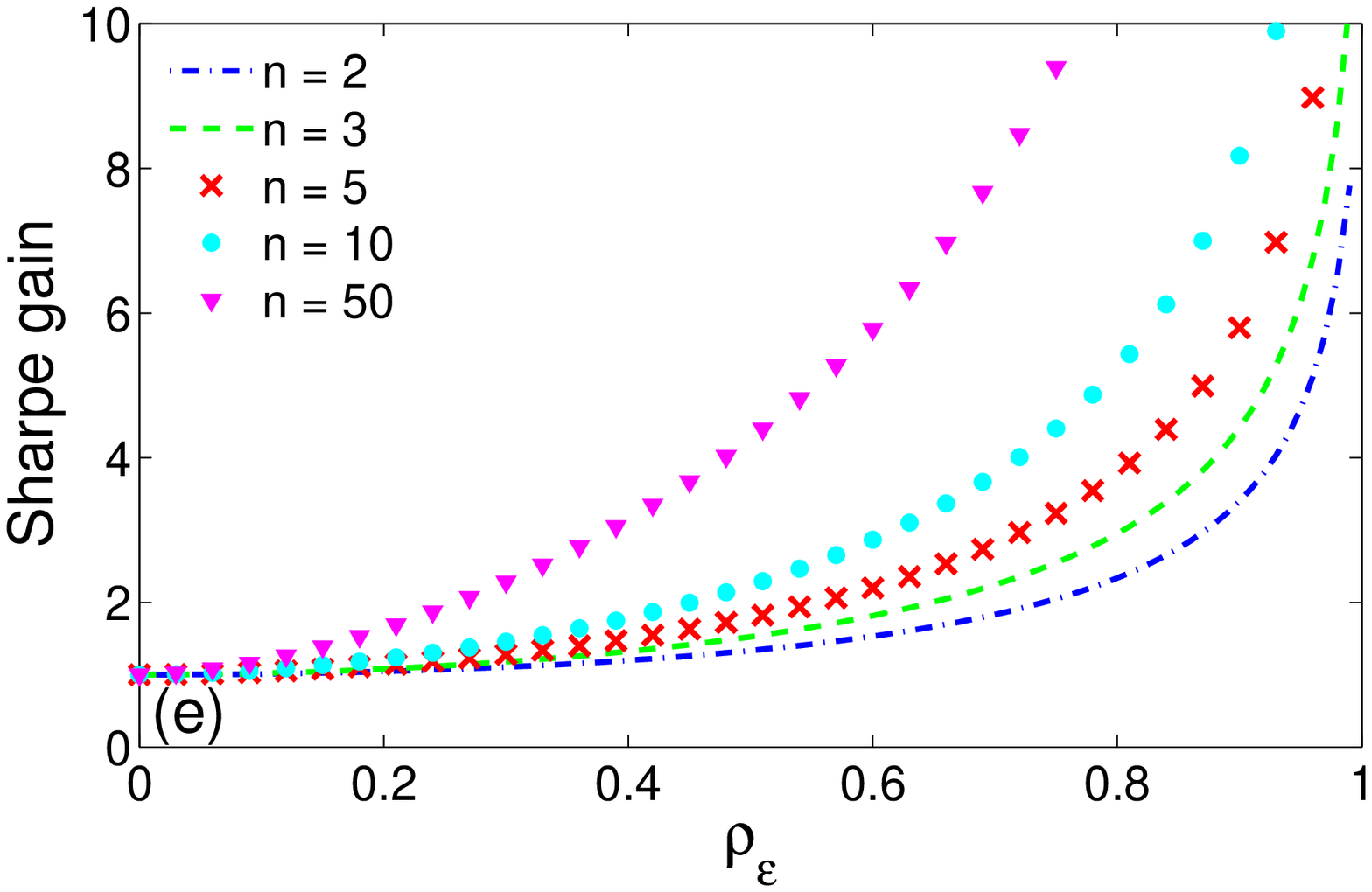}  
\includegraphics[width=0.49\textwidth]{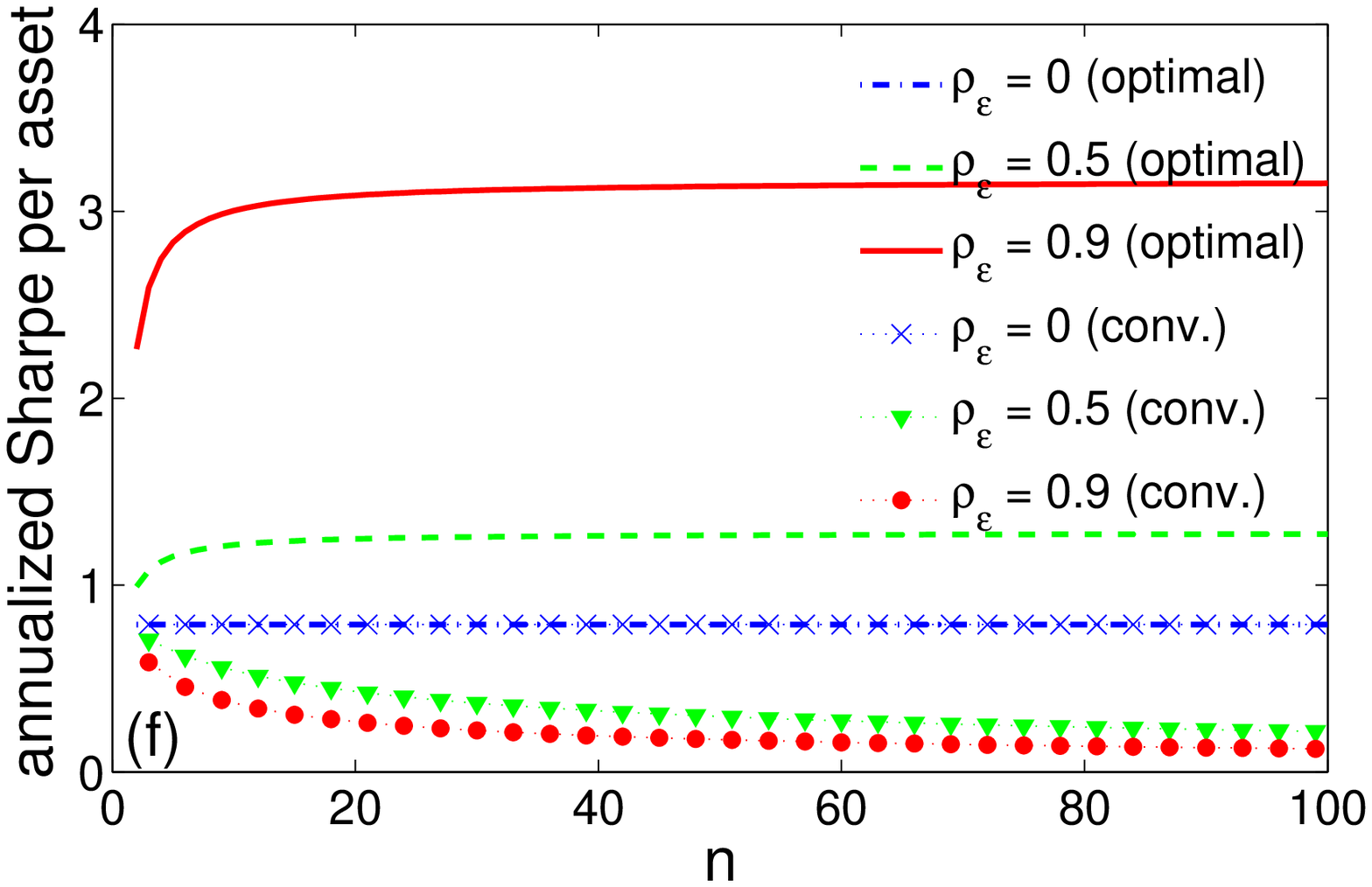}  
\end{center}
\caption{
The optimal lead-lag correction $x_{\rm opt}$ ({\bf a}), $(n-1)x_{\rm
opt}$ ({\bf b}), the annualized optimal Sharpe ratio per asset,
$\sqrt{255}~ \Sh_{\rm opt}/\sqrt{n}$ ({\bf c}), the annualized
conventional Sharpe ratio per asset, $\sqrt{255}~ \Sh_n/\sqrt{n}$
({\bf d}), and the Sharpe gain $\Sh_{\rm opt}/\Sh_n$ ({\bf e}), as
functions of $\rho_\ve$, for $n$ indistinguishable assets, with
$\rho_\xi = 0$, $\beta^j_0 = 0.1$, $\lambda = \eta = 0.01$.  The last
plot ({\bf f}) shows the annualized optimal Sharpe ratio per asset
$\sqrt{255}~ \Sh_{\rm opt}/\sqrt{n}$ (lines) and the annualized
conventional Sharpe ratio per asset $\sqrt{255}~ \Sh_n/\sqrt{n}$
(symbols) as functions of $n$.}
\label{fig:multi_rhoxi0}
\end{figure}

When $\rho_\xi = 0$, Eqs. (\ref{eq:xopt_multi}, \ref{eq:Sopt_multi})
yield
\begin{equation}
x_{\rm opt} = - \frac{V_2}{V_3} , \qquad \Sh^2_{\rm opt} = n q^2 (1-p^2) \frac{V_3}{V_1 V_3 - V_2^2} .
\end{equation}
Figure \ref{fig:multi_rhoxi0} illustrates the dependence of the
lead-lag correction $x_{\rm opt}$ and the annualized optimal Sharpe
ratio $\sqrt{255}~ \Sh_{\rm opt}$ on $\rho_\ve$.  As expected,
negative optimal lead-lag corrections are needed to reduce positive
noise correlations.  Larger $n$ require smaller correction amplitude
$|x_{\rm opt}|$ (Fig. \ref{fig:multi_rhoxi0}a).  At the same time,
each asset has $n-1$ identical lead-lag corrections (from other $n-1$
assets) so that the total correction appears as $(n-1) x_{\rm opt}$
(Fig. \ref{fig:multi_rhoxi0}b).  For large $n$, the total correction
rapidly reaches the level $-1$, even for relatively small $\rho_\ve$.
In other words, when a large number of assets is traded, even small
inter-asset correlations, if ignored, can significantly reduce the
Sharpe ratio.  One needs therefore to include lead-lag corrections.
In order to understand the rapid approach to the limiting value $-1$,
one can expand $(n-1)x_{\rm opt}$ in terms of a small parameter
$1/(n-1)$ in the limit of large $n$ as
\begin{equation}
(n-1) x_{\rm opt} \simeq -1 + \frac{Q^2(1-\rho_\ve)^2 + 2Q(1-\rho_\ve) + R}{Q^2 \rho_\ve^2~(n-1)^2}  + O\left(\frac{1}{(n-1)^3}\right) .
\end{equation}
Note that the first correction term here is of the order of
$1/(n-1)^2$.

Figure \ref{fig:multi_rhoxi0}c shows the annualized optimal Sharpe
ratio normalized by the number of assets, $\sqrt{255}~ \Sh_{\rm
opt}/\sqrt{n}$, as a function of $\rho_\ve$.  When there is no
correlation ($\rho_\ve = 0$), this quantity does not depend on $n$, as
expected from Eq. (\ref{eq:factor_S0}), and all curves come to the
same point $\sqrt{255}~ \Sh_1 \approx 0.7885$ (for the chosen set of
parameters).  When $\rho_\ve$ increases, the annualized optimal Sharpe
ratio per asset also monotonously increases.  This is in sharp
contrast to the annualized Sharpe ratio per asset for conventional
trading without lead-lag correction, $\sqrt{255}~ \Sh_n/\sqrt{n}$,
shown on Fig. \ref{fig:multi_rhoxi0}d.  As expected, inter-asset
correlations reduce diversification and thus diminish $\Sh_n/\sqrt{n}$
if lead-lag terms are ignored.  Comparison of
Figs. \ref{fig:multi_rhoxi0}c and
\ref{fig:multi_rhoxi0}d suggests that inter-asset correlations can
significantly increase the Sharpe ratio by inclusion of the lead-lag
terms.  In contrast to conventional views, these correlations, if
correctly accounted for, are not deteriorative but beneficial.  Figure
\ref{fig:multi_rhoxi0}e shows the Sharpe gain (i.e., the ratio between
the optimal and conventional Sharpe ratios, $\Sh_{\rm opt}/\Sh_n$) due
to accounting for lead-lag corrections.  This effect is particularly
important for large $\rho_\ve$ and large $n$.  It is also worth noting
the difference with the earlier case $\rho_\ve = \rho_\xi$, for which
the annualized optimal Sharpe ratio per asset was independent of both
$n$ and correlations.  In other words, equal inter-asset correlations
between noises and stochastic trends do not provide opportunities for
increasing the Sharpe ratio with correlations.  In turn, correlations
only between noises allow to TF strategies to better estimate and then
eliminate their effects, enhancing contributions from stochastic
trends.

Interestingly, the curves on Fig. \ref{fig:multi_rhoxi0}c for
different $n$ do not coincide, as one might expect from the
uncorrelated case.  The larger the number of assets $n$, the faster
increase of $\sqrt{255}~ \Sh_{\rm opt}/\sqrt{n}$.  In other words, the
Sharpe ratio of the optimal portfolio grows slightly {\it faster} than
$\sqrt{n}$.  At the same time, these curves progressively approach to
the limiting curve as $n\to\infty$
\begin{equation}
\lim\limits_{n\to\infty} \frac{\Sh^2_{\rm opt}}{n} =  \frac{q^2 (1-p^2)}{Q^2(1-\rho_\ve)^2 + 2Q(1-\rho_\ve) + R} .
\end{equation}
This explicit function (shown by black solid line) accurately
reproduces the annualized optimal Sharpe ratio per asset for moderate
$n = 50$.  A rapid approach to the limit is illustrated on
Fig. \ref{fig:multi_rhoxi0}f which shows (by lines) the annualized
optimal Sharpe ratio per asset, $\sqrt{255}~ \Sh_{\rm opt}/\sqrt{n}$,
as a function of $n$.  This quantity rapidly saturates to a constant
level, in contrast to $\sqrt{255}~ \Sh_n/\sqrt{n}$ for conventional
trading which progressively diminishes with $n$ (shown by symbols).

\subsection{Uncorrelated noises ($\rho_\ve = 0$)}

\begin{figure}
\begin{center}
\includegraphics[width=0.49\textwidth]{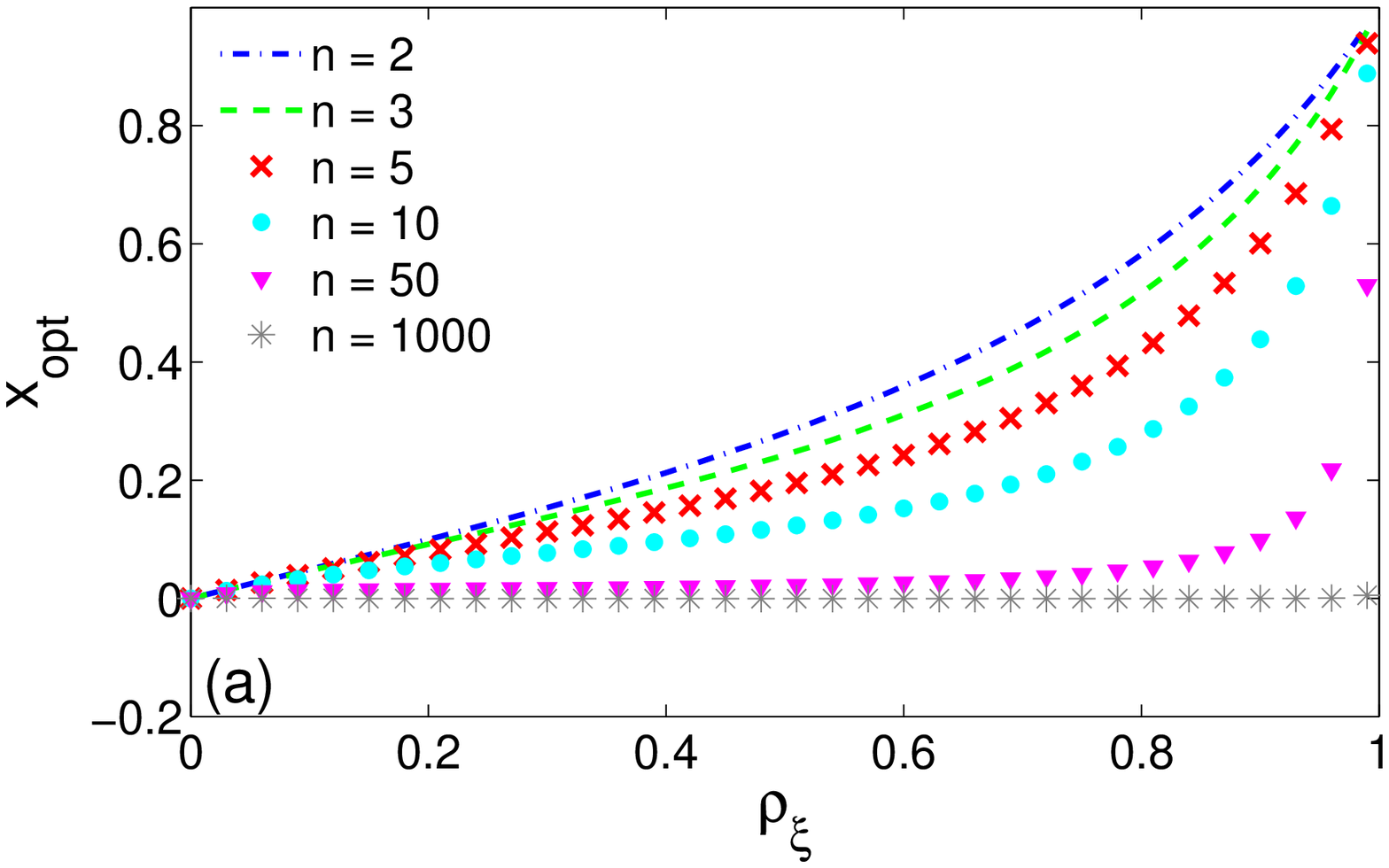} 
\includegraphics[width=0.49\textwidth]{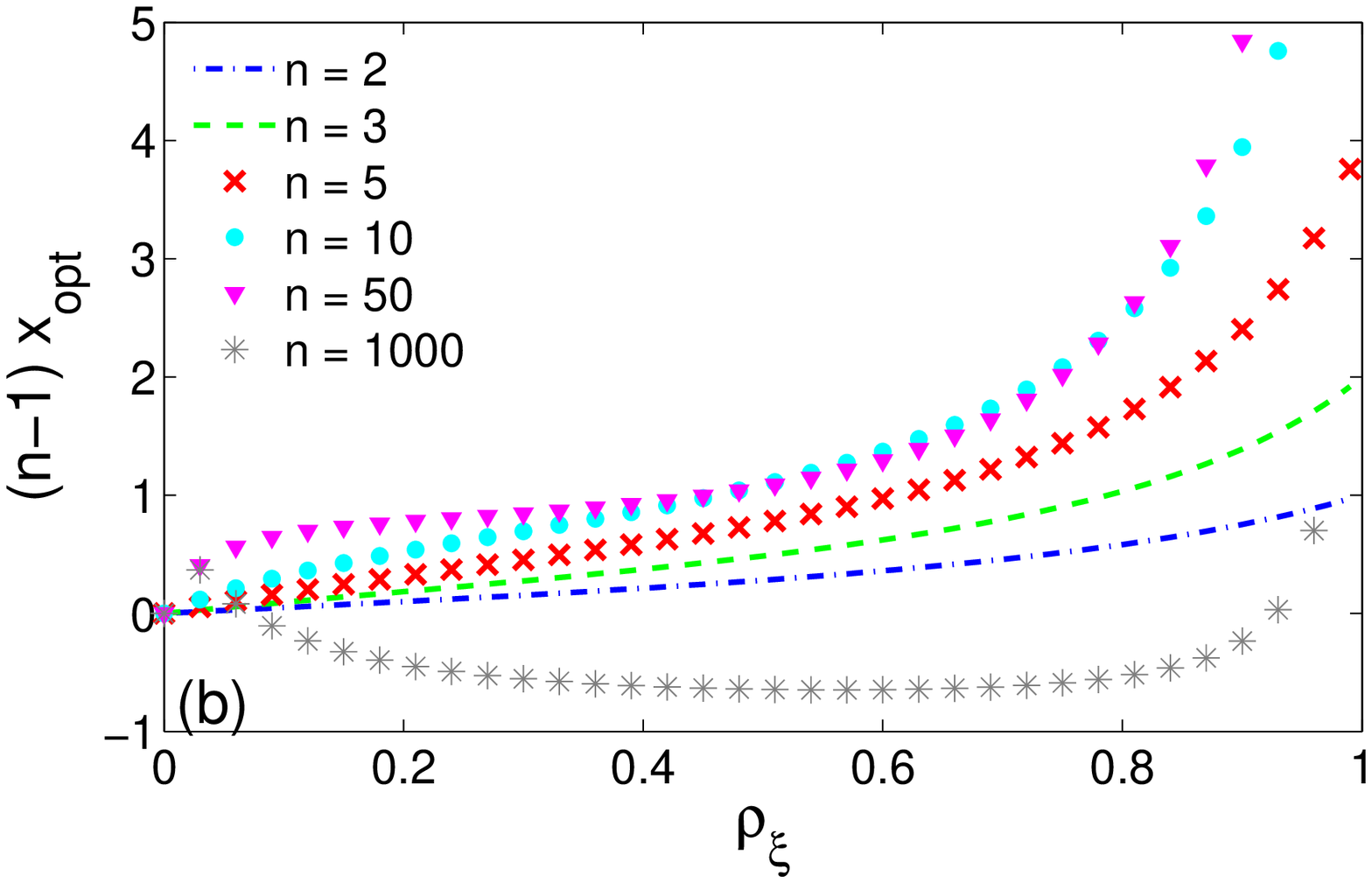} 
\includegraphics[width=0.49\textwidth]{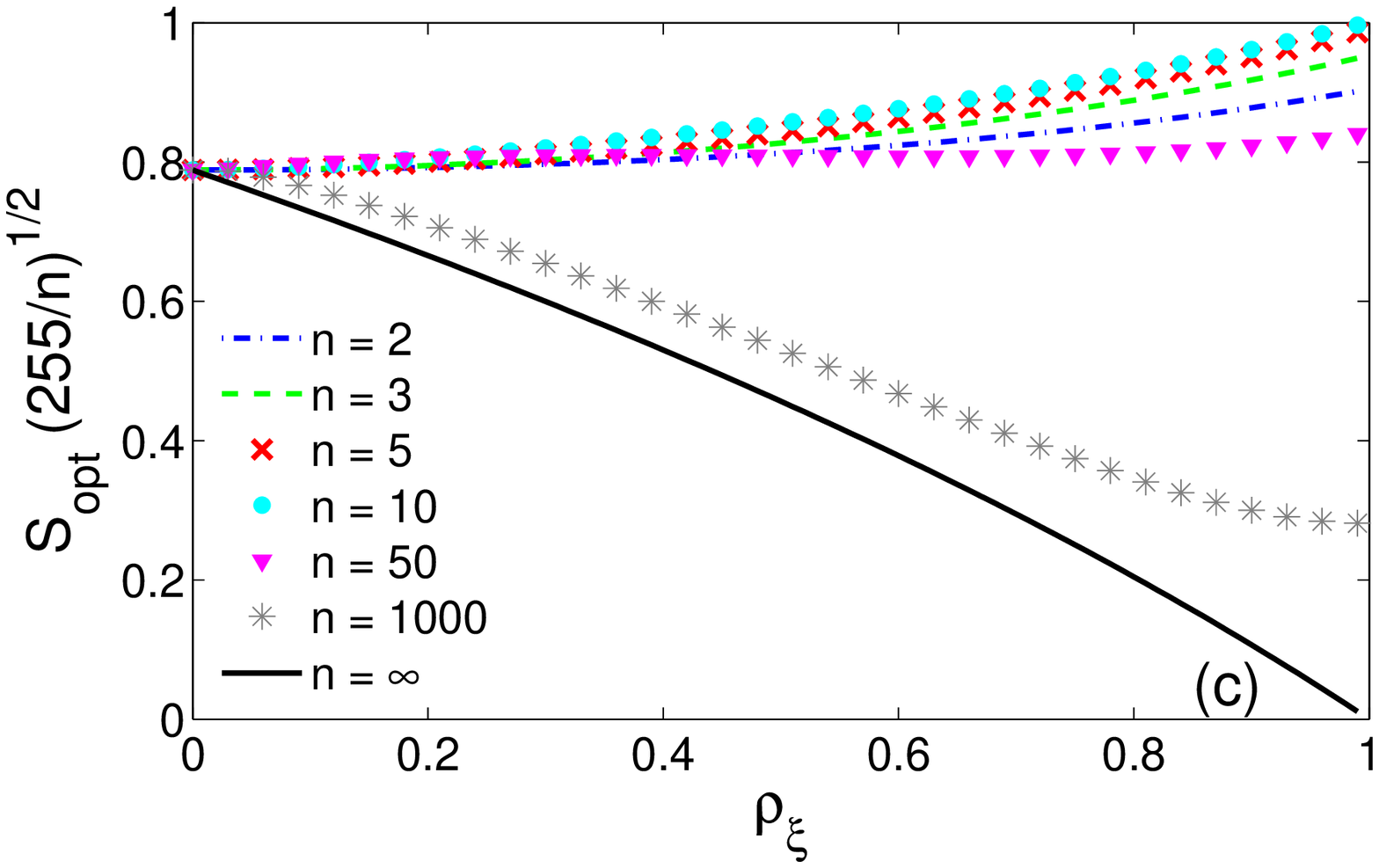} 
\includegraphics[width=0.49\textwidth]{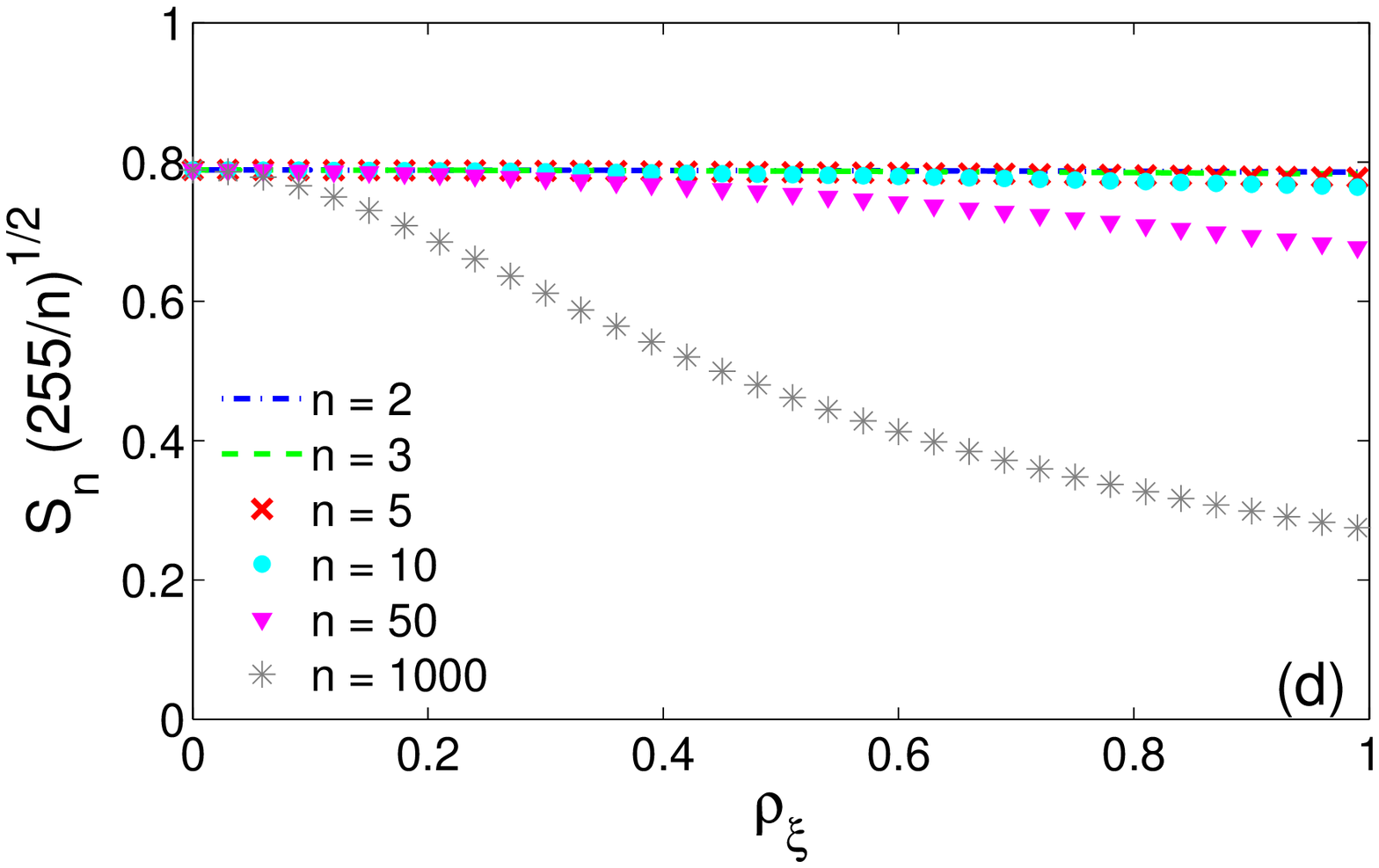} 
\includegraphics[width=0.49\textwidth]{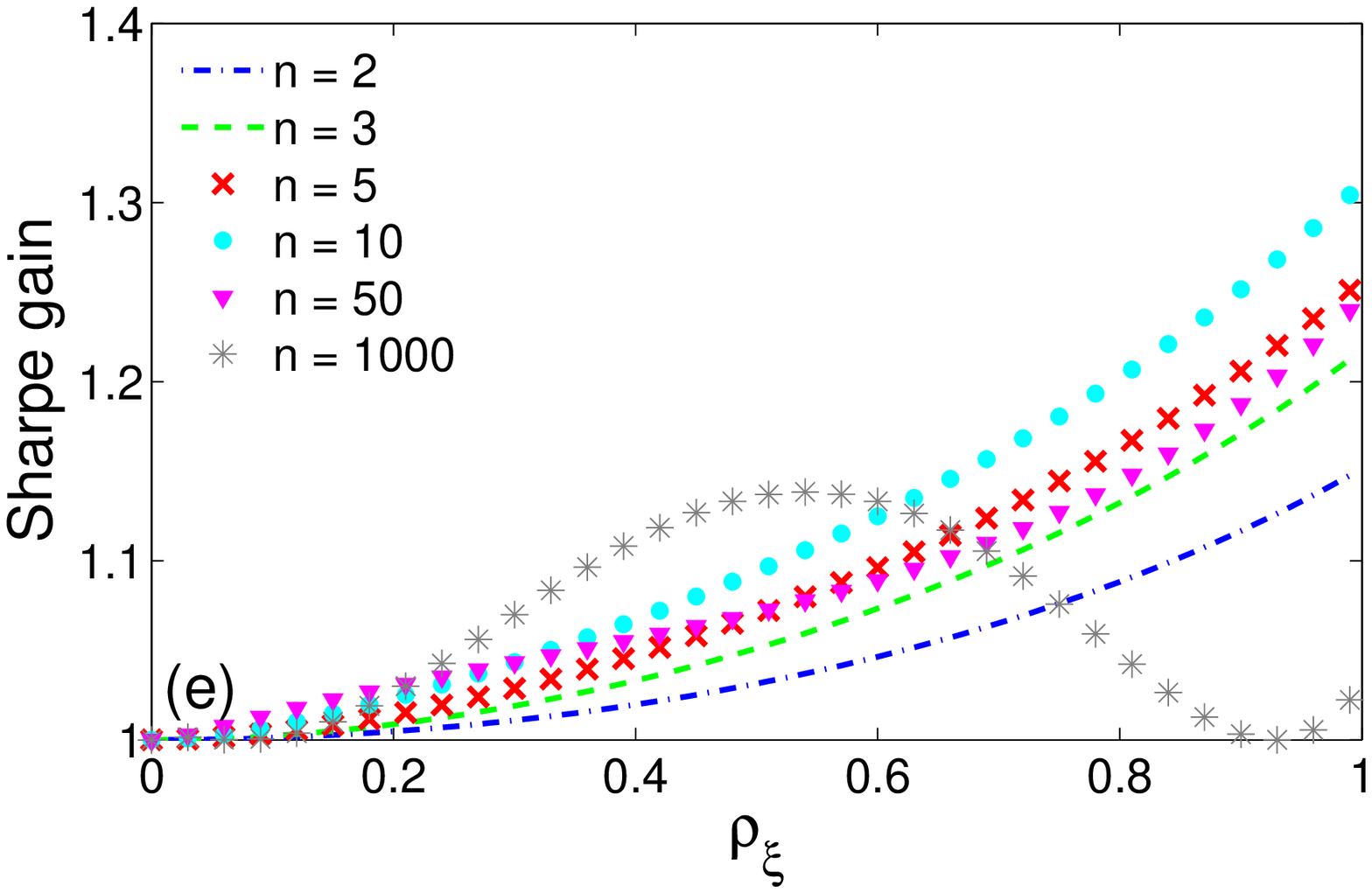} 
\includegraphics[width=0.49\textwidth]{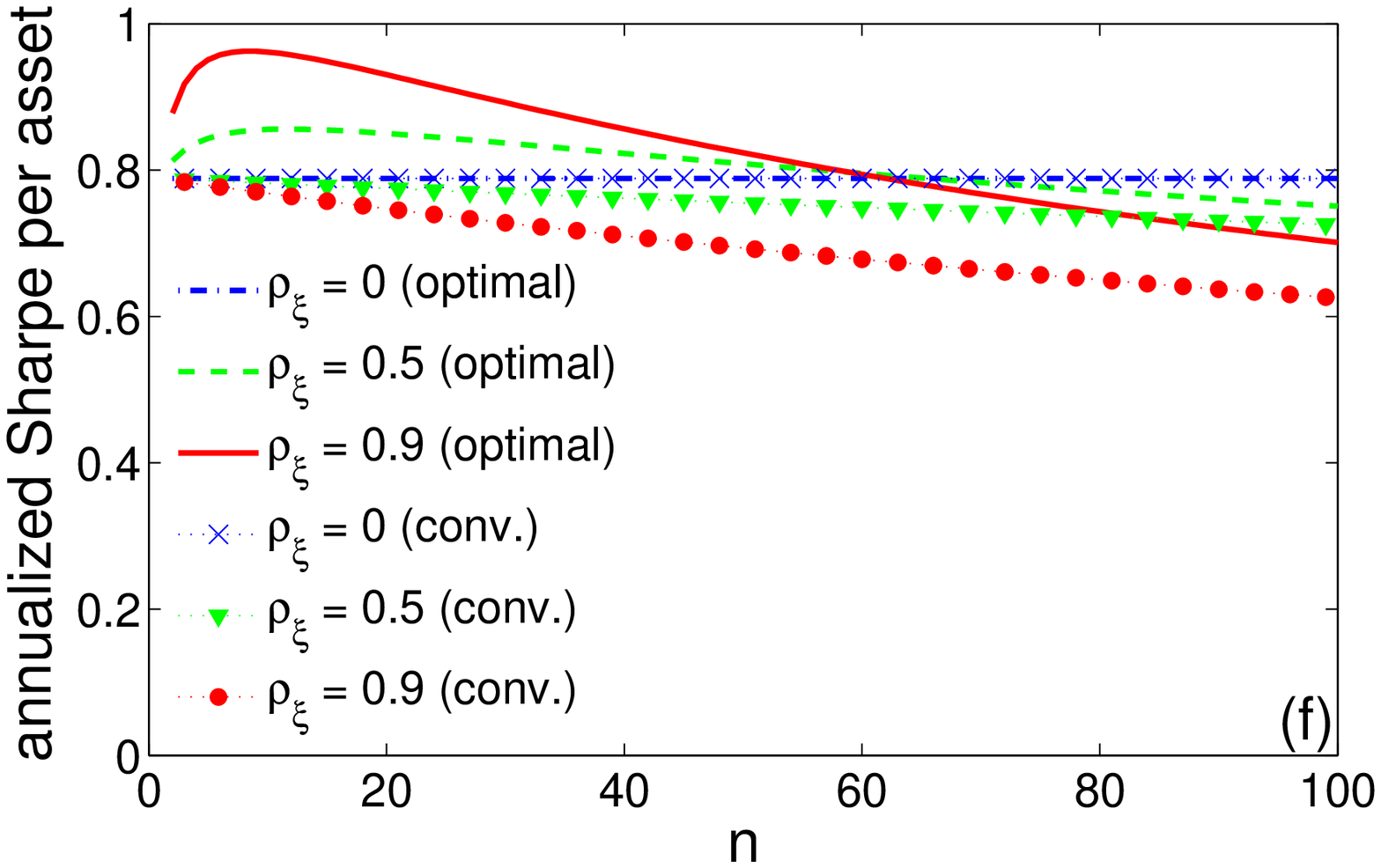} 
\end{center}
\caption{
The optimal lead-lag correction $x_{\rm opt}$ ({\bf a}), $(n-1)x_{\rm
opt}$ ({\bf b}), the annualized optimal Sharpe ratio per asset,
$\sqrt{255}~ \Sh_{\rm opt}/\sqrt{n}$ ({\bf c}), the annualized
conventional Sharpe ratio per asset, $\sqrt{255}~ \Sh_n/\sqrt{n}$
({\bf d}), and the Sharpe gain $\Sh_{\rm opt}/\Sh_n$ ({\bf e}), as
functions of $\rho_\xi$, for $n$ indistinguishable assets, with
$\rho_\ve = 0$, $\beta^j_0 = 0.1$, $\lambda = \eta = 0.01$.  The last
plot ({\bf f}) shows the annualized optimal Sharpe ratio per asset
$\sqrt{255}~ \Sh_{\rm opt}/\sqrt{n}$ (lines) and the annualized
conventional Sharpe ratio per asset $\sqrt{255}~ \Sh_n/\sqrt{n}$
(symbols) as functions of $n$.}
\label{fig:multi_rhoeps0}
\end{figure}

The other limiting case $\rho_\ve = 0$ is even more intriguing.
Figure \ref{fig:multi_rhoeps0} shows the optimal lead-lag correction
$x_{\rm opt}$, the annualized optimal Sharpe ratio per asset,
$\sqrt{255}~ \Sh_{\rm opt}/\sqrt{n}$, the annualized conventional
Sharpe ratio per asset $\sqrt{255}~ \Sh_n/\sqrt{n}$, and the Sharpe
gain $\Sh_{\rm opt}/\Sh_n$, as functions of $\rho_\xi$.

The optimal lead-lag correction $x_{\rm opt}$ is positive
(Fig. \ref{fig:multi_rhoeps0}a), as for the two-asset case.  This
observation may sound counter-intuitive because one might expect that
a negative lead-lag term is needed to correct for positively
correlated stochastic trends.  Moreover, once $x_{\rm opt}$ is
multiplied by $n-1$ (Fig. \ref{fig:multi_rhoeps0}b), one can clearly
see that the lead-lag term is indeed negative for $n = 1000$ (but
still remaining positive for small and moderate $n$).  Most
surprisingly, the lead-lag correction for $n = 1000$ is positive for
small and large $\rho_\xi$, while negative for intermediate values.
In order to clarify this situation, we consider the asymptotic
behavior of the optimal lead-lag correction for large $n$:
\begin{equation}
(n-1) x_{\rm opt}\simeq - 1 + \frac{Q^2 + 2Q(1-\rho_\xi) + R(1-\rho_\xi)^2}{\rho_\xi(1-\rho_\xi) R~ (n-1)} + O\left(\frac{1}{(n-1)^2}\right) .
\end{equation}
As intuitively expected, the total lead-lag correction approaches to
$-1$, as for the earlier case $\rho_\xi = 0$ from
Sec. \ref{sec:factor_noise}.  Although the first-order correction term
vanishes in the limit $n\to\infty$, the coefficient in front of this
term is large because $R \ll Q$.  As a consequence, one needs to
consider thousands of assets in order to approach the limit $-1$, in
sharp contrast to the earlier case from Sec. \ref{sec:factor_noise}.
We conclude that, for the present case $\rho_\ve = 0$, the asymptotic
formulas in the limit $n\to\infty$ are not useful, and may even be
misleading when applied to moderate number of assets.  In particular,
we found positive lead-lag correction for the two-asset case in
Sec. \ref{sec:Stwo_indist}.

Another surprising feature appears in the non-monotonous behavior of
the annualized optimal Sharpe ratio per asset, $\sqrt{255}~ \Sh_{\rm
opt}/\sqrt{n}$ (Fig. \ref{fig:multi_rhoeps0}c).  One can see that this
quantity monotonously grows with $\rho_\xi$ for small number of assets
(up to 10), exhibits non-monotonous behavior for moderate $n = 50$,
and decreases for very large number of assets ($n = 1000$).  Moreover,
the curve for $n = 10$ lies above the curve for $n = 50$, in sharp
contrast to Fig. \ref{fig:multi_rhoxi0}c.  This is also seen on
Fig. \ref{fig:multi_rhoxi0}f which shows the same quantity as a
function of $n$.  Note, however, that the total annualized Sharpe
ratio, $\sqrt{255}~ \Sh_{\rm opt}$, is still higher for larger $n$,
due to the factor $\sqrt{n}$.  In other words, if adding more assets
with correlated noises ($\rho_\ve > 0$) increased the Sharpe ratio
even more than by factor $\sqrt{n}$, adding more assets with
correlated stochastic trends ($\rho_\xi > 0$) may increase the Sharpe
ratio by less than the factor $\sqrt{n}$.  This is particularly clear
from the comparison of the Sharpe gain on Fig. \ref{fig:multi_rhoxi0}e
and \ref{fig:multi_rhoeps0}e: for correlated stochastic trends
($\rho_\xi> 0$), the Sharpe gain is modest (up to $20\%$) even at very
high correlation coefficient $\rho_\xi$; in contrast, the Sharpe gain
was extremely large (a factor 10 or higher) for correlated noises
($\rho_\ve > 0$).  The decrease of the annualized optimal Sharpe ratio
per asset for $n = 1000$ can be understood by considering the limit
$n\to\infty$, for which
\begin{equation}
\lim\limits_{n\to\infty} \frac{\Sh^2_{\rm opt}}{n} = q^2 (1-p^2) \frac{(1-\rho_\xi)^2}{Q^2 + 2Q(1-\rho_\xi) + R(1-\rho_\xi)^2} .
\end{equation}
This limiting function (shown by black solid line on
Fig. \ref{fig:multi_rhoeps0}c) monotonously decreases with $\rho_\xi$.
As for $x_{\rm opt}$, the approach to the limit is very slow so that
the limiting curve does not capture the behavior for moderately large
$n$.  We conclude that the case of correlated stochastic trends
exhibits some counter-intuitive features that require particular
attention from fund managers.

\section*{Conclusion}
\label{sec:conclusion}

The principle of diversification in portfolio management
\cite{Sharpe,Ilmanen12} calls for investing in as many 
uncorrelated assets as possible in order to reduce a portfolio risk.
The same principle is applied to trend following portfolios.  In a
typical setting, the exposure of each TF strategy is determined by
earlier returns of the traded asset, independently of other assets.
Then, the weight of each strategy is adjusted to account for asset
covariance structure.  In this setting, investing in correlated assets
is more risky and thus yields smaller risk-adjusted returns (or Sharpe
ratio).  However, properly modeling the source of correlations can be
beneficial as a mean to estimate apparent trends more reliably, to
adjust the TF portfolio more efficiently, and thus to enhance the
Sharpe ratio.  The paper aimed at analyzing this fact in a
quantitative way.  For this purpose, we introduced a simple Gaussian
model, in which volatility-normalized returns of each asset had two
contributions: short-range noises mimicking instantaneous {\it
uninformative} price fluctuations (e.g., daily), and stochastic trends
modeling TF profitability on a longer time scale (e.g., monthly).

The Gaussian assumption allowed us to derive analytically the mean and
variance of the portfolio profit-and-loss and thus to formulate
explicitly the problem of Sharpe ratio maximization.  This problem was
solved analytically for two assets and for a sector model of $n$
similar assets, while an exact numerical solution is possible in
general.

As mentioned earlier, each strategy should incorporate information
from other strategies.  We considered each asset investment as a
linear combination of all assets strategy signals.  The weight
$\omega_{j,k}$ represents a correction term to the $j$-th asset
investment due to its correlation to the $k$-th asset.  This cross
correcting, or lead-lag, term is indeed proportional to the $k$-th
strategy signal, itself being a linear combination of $k$-th asset
past returns.  We compared optimal portfolios formed by these
augmented strategies with optimal portfolios of individual strategies.
Instead of determining the weights of $n$ TF strategies (one for each
asset), our generalized portfolio allocation problem operates with
$n^2$ weights $\omega_{j,k}$.

One may look at it as a static allocation problem on $n^2$ virtual
strategies, which include the $n$ conventional individual trading
strategies, plus $n(n-1)$ lead-lag strategies, contributing to an
asset position using past price variations of another one.  We showed
that when there is no correlation between assets, both approach are
equivalent and the optimal weight matrix $\omega_{j,k}$ is diagonal.
However, in presence of inter-asset correlations, optimal portfolio
weights contain non-diagonal terms.  As a consequence, the static
allocation scheme is sub-optimal while the introduction of the
lead-lag terms with $\omega_{j,k}$ made inter-asset correlations
highly beneficial, as we demonstrated on a simple two-asset portfolio.

We also investigated the respective roles of noise-noise ($\rho_\ve$)
and trend-trend ($\rho_\xi$) inter-asset correlations.  The separate
accounting for these two mechanisms is a new feature of our model.
When $\rho_\ve = \rho_\xi$ (i.e., the same structure of inter-asset
correlations for noises and trends), the optimal Sharpe ratio for $n$
indistinguishable assets is equal to $\sqrt{n}$ times the Sharpe ratio
of one asset, as expected due to diversification.  The inclusion of
the lead-lag corrections $\omega_{j,k}$ allows one to reach this
optimal Sharpe ratio even for strongly correlated assets but one
cannot overperform here the benchmark case of uncorrelated assets.  In
turn, when $\rho_\ve \ne \rho_\xi$, the optimal Sharpe ratio can be
much larger than that for the benchmark case due to statistically more
reliable estimations.  Since economical and financial mechanisms
behind short-range price fluctuations and longer trends may be quite
different, one can speculate about hidden opportunities due to
possible mismatches between $\rho_\ve$ and $\rho_\xi$.  Perhaps,
numerous algorithmic tricks and empirical hints in trend following
strategies, as well as managers' intuition, aim to catch these
opportunities in practice.  Our study presents a first step towards
better understanding of these mechanisms.  Although the Gaussian model
remains simplistic, while statistical calibration of its parameters
from financial time series is challenging, the generalized portfolio
allocation problem is a promising way for trend followers to make
inter-asset correlations profitable.


\appendix
\section{Mean and variance of incremental P\&L of a portfolio}
\label{sec:mean_var}

From Eq. (\ref{eq:dPNL_general}), the mean incremental profit-and-loss
of a portfolio is simply
\begin{equation}
\label{eq:A_PNL_mean}
\begin{split}
\langle \dPNL_t \rangle & = \sum\limits_{j,k=1}^n  \omega_{j,k} \sum\limits_{t'=1}^{t-1} \S^k_{t,t'} \langle r_t^j r_{t'}^k \rangle 
 = \sum\limits_{j,k=1}^n  \omega_{j,k} \C_{\xi}^{j,k} (\S^k \A^k\A^{j,\T})_{t,t} , \\
\end{split}
\end{equation}
where the explicit structure (\ref{eq:C}) of the covariance $\langle
r_t^j r_{t'}^k \rangle$ was used.

Next, we compute the variance of $\dPNL_t$ as
\begin{equation}
\begin{split}
\var\{ \dPNL_t \} & = \sum\limits_{j_1,k_1=1}^n  \omega_{j_1,k_1} \sum\limits_{t'_1=1}^{t-1} (\S^{k_1})_{t,t'_1} 
\sum\limits_{j_2,k_2=1}^n  \omega_{j_2,k_2} \sum\limits_{t'_2=1}^{t-1} (\S^{k_2})_{t,t'_2} \\
& \times \biggl[\langle r_t^{j_1} r_{t}^{j_2} \rangle \langle r_{t'_1}^{k_1} r_{t'_2}^{k_2} \rangle + 
\langle r_t^{j_1} r_{t'_2}^{k_2} \rangle \langle r_t^{j_2} r_{t'_1}^{k_1} \rangle\biggr] ,   \\
\end{split}
\end{equation}
where we used the Wick's theorem for Gaussian returns $r_t^k$.
Substituting again the covariances from Eq. (\ref{eq:C}), one gets
\begin{equation*}
\begin{split}
\var\{ \dPNL_t \} & = \sum\limits_{j_1,k_1=1}^n  \omega_{j_1,k_1} \sum\limits_{t'_1=1}^{t-1} (\S^{k_1})_{t,t'_1} 
\sum\limits_{j_2,k_2=1}^n  \omega_{j_2,k_2} \sum\limits_{t'_2=1}^{t-1} (\S^{k_2})_{t,t'_2} \\
& \times \biggl[\bigl(\C_\ve^{j_1,j_2} + \C_{\xi}^{j_1,j_2} (\A^{j_1}\A^{j_2,\T})_{t,t}\bigr)
\bigl(\delta_{t'_1,t'_2}\C_\ve^{k_1,k_2} + \C_{\xi}^{k_1,k_2} (\A^{k_1}\A^{k_2,\T})_{t'_1,t'_2}\bigr) \\
& + \bigl(\C_{\xi}^{j_1,k_2} (\A^{j_1}\A^{k_2,\T})_{t,t'_2}\bigr)
\bigl(\C_{\xi}^{j_2,k_1} (\A^{j_2}\A^{k_1,\T})_{t,t'_1}\bigr) \biggr] ,  \\
\end{split}
\end{equation*}
from which
\begin{equation}
\label{eq:A_PNL_var}
\begin{split}
\var\{ \dPNL_t \} & = \sum\limits_{j_1,k_1,j_2,k_2=1}^n  \omega_{j_1,k_1} \omega_{j_2,k_2}  
\biggl[\bigl(\C_\ve^{j_1,j_2} + \C_\xi^{j_1,j_2} (\A^{j_1}\A^{j_2,\T})_{t,t}\bigr) \\
& \times \bigl(\C_\ve^{k_1,k_2} (\S^{k_1} \S^{k_2,\T})_{t,t} + \C_{\xi}^{k_1,k_2} 
(\S^{k_1} \A^{k_1} \A^{k_2,^\T} \S^{k_2,\T})_{t,t} \bigr)  \\
& + \C_{\xi}^{j_1,k_2} \C_{\xi}^{k_1,j_2} (\S^{k_1} \A^{k_1,\T} \A^{j_2})_{t,t}  
(\S^{k_2} \A^{j_1,\T} \A^{k_2})_{t,t}  \biggr] . \\
\end{split}
\end{equation}
This is the variance of the incremental P\&L in the general case.
Note that Eqs. (\ref{eq:A_PNL_mean}, \ref{eq:A_PNL_var}) can be
re-written in the form (\ref{eq:mean_var}, \ref{eq:MV_general}).

In what follows, we introduce a simplifying assumption that all the
assets have the same rates: $\lambda_j = \lambda$ (i.e., $q_j = q$).
Similarly, we assume that all the trend following strategies have the
same rate: $\eta_k = \eta$ (i.e., $p_k = p$).  In this case, the
elements $M_t^{j,k}$ and $V_t^{j_1,k_1; j_2,k_2}$ from
Eq. (\ref{eq:MV_general}) become
\begin{equation}
\begin{split}
M_t^{j,k} & = \gamma^k (\E_p \E_q \E_q^\T)_{t,t}  \C_{\xi,\beta}^{j,k} , \\
V_t^{j_1,k_1; j_2,k_2} & = \gamma^{k_1} \gamma^{k_2} \biggl[\C_\ve^{j_1,j_2} \C_\ve^{k_1,k_2} (\E_p \E_p^\T)_{t,t} + \C_\ve^{k_1,k_2} 
\C_{\xi,\beta}^{j_1,j_2} (\E_p \E_p^\T)_{t,t} (\E_q\E_q^\T)_{t,t} \\
& + \C_\ve^{j_1,j_2} \C_{\xi,\beta}^{k_1,k_2} (\E_p \E_q \E_q^\T \E_p^\T)_{t,t} 
+ \C_{\xi,\beta}^{j_1,j_2} \C_{\xi,\beta}^{k_1,k_2} (\E_q \E_q^\T)_{t,t} (\E_p \E_q \E_q^\T \E_p^\T)_{t,t}   \\
& + \C_{\xi,\beta}^{j_1,k_2} \C_{\xi,\beta}^{k_1,j_2} [(\E_p \E_q^\T \E_q)_{t,t}]^2 \biggr],
\end{split}
\end{equation}
where $[\E_q]_{t,t'} = q^{t-t'-1}$ for $t>t'$, and $0$ otherwise, and
$\C_{\xi,\beta}^{j,k} \equiv \beta^j \beta^k \C_{\xi}^{j,k}$.
Supplementary Materials to \cite{Grebenkov14} provide the
explicit formulas for various products of matrices $\E_p$ and $\E_q$.
In the stationary limit $t\to\infty$, one gets
\begin{equation}
\begin{split}
\lim\limits_{t\to\infty} (\E_p \E_p^\T)_{t,t} & = \frac{1}{1-p^2} , \\
\lim\limits_{t\to\infty} (\E_p \E_q \E_q^\T)_{t,t} & = \frac{q}{(1-q^2)(1-pq)} , \\
\lim\limits_{t\to\infty} (\E_p \E_q \E_q^\T \E_p^\T)_{t,t} & = \frac{1+pq}{(1-pq)(1-q^2)(1-p^2)} , \\
\end{split}
\end{equation}
from which we obtain
\begin{equation}
\label{eq:M_V_st}
\begin{split}
M_\infty^{j,k} & = \frac{q \sqrt{1-p^2}}{1-pq}~ \C_{\xi,\beta_0}^{j,k} , \\
V_\infty^{j_1,k_1; j_2,k_2} & = \C_\ve^{j_1,j_2} \C_\ve^{k_1,k_2} + \C_\ve^{k_1,k_2} 
\C_{\xi,\beta_0}^{j_1,j_2} + \C_\ve^{j_1,j_2} \C_{\xi,\beta_0}^{k_1,k_2} \frac{1+pq}{1-pq} \\
& + \C_{\xi,\beta_0}^{j_1,j_2} \C_{\xi,\beta_0}^{k_1,k_2} \frac{1+pq}{1-pq}   
 + \C_{\xi,\beta_0}^{j_1,k_2} \C_{\xi,\beta_0}^{k_1,j_2} \frac{q^2(1-p^2)}{(1-pq)^2} ,  \\
\end{split}
\end{equation}
where $\C_{\xi,\beta_0}^{j,k} \equiv \C_{\xi,\beta}^{j,k}/(1-q^2)$,
and we set $\gamma^k = \sqrt{1 - (1-\eta^k)^2} = \sqrt{1-p^2}$.  This
normalization was proposed in \cite{Grebenkov14} to set the unit
variance of the stationary incremental P\&L of a single asset without
auto-correlations (i.e., when $\beta = 0$).  For the multivariate
case, this normalization yields the expected form $V_\infty^{j_1,k_1;
j_2,k_2} = \C_\ve^{j_1,j_2} \C_\ve^{k_1,k_2}$ when all $\beta^j = 0$.
For symmetric weights, the above expression for the covariance matrix
can be further simplified to get Eq. (\ref{eq:M_V_st2}).

\section{Two indistinguishable assets}
\label{sec:two_indist}

When $\beta^1_0 = \beta^2_0 = \beta_0$ (i.e., $\kappa = 1$) and
$\sigma^1 = \sigma^2 = 1$ (i.e., $\nu = 1$), two assets have the same
structure of auto-correlations that makes them indistinguishable from
each other.  In this case, Eqs. (\ref{eq:mean_var_two},
\ref{eq:Omega}) are reduced to
\begin{equation}
\Sh^2 = q^2 (1-p^2) \frac{(\omega_{11} + 2\rho_\xi \omega_{12} + \omega_{22})^2}{Q^2\Omega_1 + 2Q\Omega_2 + R\Omega_3}  , 
\end{equation}
with
\begin{equation}
\label{eq:Omegas_1}
\begin{split}
\Omega_1 & =  \omega_{11}^2 + 2\omega_{12}^2 + \omega_{22}^2 + 2\rho_\ve^2(\omega_{12}^2 + \omega_{11}\omega_{22})
+ 4\rho_\ve \omega_{12}(\omega_{11} + \omega_{22}) , \\
\Omega_2 & = \omega_{11}^2 + 2\omega_{12}^2 + \omega_{22}^2 + 2\rho_\ve \rho_\xi (\omega_{12}^2+\omega_{11}\omega_{22})
+ 2(\rho_\ve + \rho_\xi) \omega_{12}(\omega_{11} + \omega_{22}) , \\
\Omega_3 & = \omega_{11}^2 + 2\omega_{12}^2 + \omega_{22}^2 + 2\rho_\xi^2(\omega_{12}^2 + \omega_{11}\omega_{22})
+ 4\rho_\xi \omega_{12}(\omega_{11} +\omega_{22}) .  \\
\end{split}
\end{equation}
In this case, three quadratic equations determining the weights
ratios, $z = \omega_{11}/\omega_{22}$ and $x =
\omega_{12}/\omega_{22}$, are
\begin{equation}
\label{eq:auxil_eqn}
\begin{split}
2Ax^2 + 2Bxz + 2Cx - Dz + D &= 0 , \\
Dz^2 + 2Ax^2 + 2Cxz - Dz + 2Bx &= 0, \\
Bz^2 + Axz + 2Cz + Ax + B &= 0, \\
\end{split}
\end{equation}
where
\begin{equation}
\begin{split}
A &= Q^2(1-2\rho_\ve \rho_\xi + \rho_\ve^2) + 2Q(1-\rho_\xi^2) + R(1-\rho_\xi^2), \\
B &= Q(Q+1)(\rho_\ve - \rho_\xi), \\
C &= Q^2\rho_\ve (1-\rho_\ve \rho_\xi) + Q(\rho_\ve + \rho_\xi - 2\rho_\ve \rho_\xi^2) + R \rho_\xi(1-\rho_\xi^2) , \\
D &= Q^2(1-\rho_\ve^2) + 2Q(1-\rho_\ve \rho_\xi) + R(1-\rho_\xi^2). \\
\end{split}
\end{equation}
The difference between the first two relations in
Eqs. (\ref{eq:auxil_eqn}) yields $(z-1) [2(B-C)x - D(1+z)] = 0$, from
which one determines both $z$ and $x$.  One can show that the
quadratic equation corresponding to the choice $z = 2(B-C)x/D - 1$
does not have real solutions.  As a consequence, we get the following
solution of the minimization problem: $z_{\rm opt} = 1$ (i.e.,
$\omega_{11} = \omega_{22}$), while $x_{\rm top}$ is given by
Eq. (\ref{eq:xopt_indist}).

\section{Two assets without lead-lag term}

In the simplest situation, one can consider a linear combination of
two assets with weights $\omega_{11}$ and $\omega_{22}$, without
introducing a lead-lag term: $\omega_{12} = 0$.  In this case,
Eq. (\ref{eq:Sharpe_two}) for the squared Sharpe ratio becomes
\begin{equation}
\Sh^2 = (1-p^2)q^2 \frac{(\omega_{11} \kappa^2 + \omega_{22})^2}{a \omega_{11}^2 + 2b\omega_{11}\omega_{22} + c \omega_{22}^2} ,
\end{equation}
where
\begin{equation}
\begin{split}
a & = Q^2 + 2Q \kappa^2 + R \kappa^4 ,  \\
b & = Q^2 \rho_\ve^2 + 2Q \kappa \rho_\ve \rho_\xi + R \kappa^2 \rho_\xi^2 , \\
c & = Q^2 + 2Q + R , \\
\end{split}
\end{equation}
and we set $\sigma^1 = \sigma^2 = 1$.  The optimization leads to the
following quadratic equation on the weights
\begin{equation}
\omega_{11}^2 [b\kappa^4 - a\kappa^2] + \omega_{11}\omega_{22} [c\kappa^4-a] + \omega_{22}^2 [c\kappa^2 - b] = 0 ,
\end{equation}
whose solutions can be written explicitly:
\begin{equation}
z = \frac{\omega_{11}}{\omega_{22}} = \frac{a - c\kappa^4 \pm \sqrt{(a-c\kappa^4)^2 - 4(c\kappa^2 -b)(b\kappa^4 - a\kappa^2)}}{2(b\kappa^4 - a\kappa^2)} .
\end{equation}

In the particular case of indistinguishable assets (i.e., $\kappa =
1$), one has $a = c$, and two solutions of the above equation are
$\omega_{11} = \pm \omega_{22}$, whatever the values of $\rho_\ve$ and
$\rho_\xi$.  Note that the maximum is achieved for $\omega_{11} =
\omega_{22}$ (while $\Sh = 0$ in the opposite case $\omega_{11} =
-\omega_{22}$).  As a consequence, one needs to take the linear
combination with equal weights, as expected.  We get then $\Sh^2_0 =
\frac{2(1-p^2)q^2}{a+b}$, from which one retrieves Eq. (\ref{eq:S0}).

\section{Derivation for a sector model}
\label{sec:Afactor}

We consider the case of $n$ indistinguishable assets (with $\beta^j =
\beta$ and $\sigma^j = \sigma = 1$).  In the optimal portfolio of TF
strategies, all assets are expected to have the same weight,
$\omega_{jj} = \omega_{11}$, as well as all lead-lag corrections are
the same: $\omega_{jk} = \omega_{12}$ for all $j\ne k$.  Substituting
these weights into Eqs. (\ref{eq:mean_var}, \ref{eq:M_V_st2}), one
gets
\begin{equation}
\begin{split}
\langle \dPNL_\infty \rangle & 
= \frac{q\sqrt{1-p^2}}{1-pq} \sum\limits_{j,k}^n \C_{\xi,\beta_0}^{j,k} \omega_{j,k} 
 = \frac{q\sqrt{1-p^2} \beta_0^2}{1-pq} \bigl[\omega_{11} n + \omega_{12} n(n-1)\rho_\xi \bigr] , \\
\var\{\dPNL_\infty\} & = \sum\limits_{j_1,k_1,j_2,k_2}^n V^{j_1,k_1; j_2,k_2} \omega_{j_1,k_1} \omega_{j_2,k_2} 
 = \omega_{11}^2  \tilde{V}_1 + 2\omega_{11} \omega_{12} \tilde{V}_2 + \omega_{12}^2  \tilde{V}_3 , \\
\end{split}
\end{equation}
where
\begin{equation}
\begin{split}
\tilde{V}_1 & \equiv \sum\limits_{j,k} V^{j,j; k,k} = \sum\limits_{j} \biggl[\C_\ve^{j,k} \C_\ve^{j,k} + \frac{2\beta_0^2}{1-pq} \C_\ve^{j,k} \C_\xi^{j,k}
+ \frac{\beta_0^4 R}{(1-pq)^2} \C_\xi^{j,k} \C_\xi^{j,k}\biggr] \\
& = n \biggl[1 + \frac{2\beta_0^2}{1-pq} + \frac{\beta_0^4 R}{(1-pq)^2}\biggr] + n(n-1) \biggl[\rho_\ve^2 + \frac{2\beta_0^2}{1-pq} \rho_\ve \rho_\xi 
+ \frac{\beta_0^4 R}{(1-pq)^2} \rho_\xi^2 \biggr], \\
\tilde{V}_2 & \equiv \sum\limits_{j,j_2\ne k_2} V^{j,j; j_2,k_2} = \sum\limits_{j,j_2\ne k_2} 
\biggl[\C_\ve^{j,j_2} \C_\ve^{j,k_2} + \frac{2\beta_0^2}{1-pq} \C_\ve^{j,j_2} \C_\xi^{j,k_2} 
+ \frac{\beta_0^4 R}{(1-pq)^2} \C_\xi^{j,j_2} \C_\xi^{j,k_2}\biggr] \\
& = n(n-1) \biggl[2\rho_\ve + (n-2)\rho_\ve^2 + \frac{2\beta_0^2}{1-pq} (\rho_\ve + \rho_\xi + (n-2)\rho_\ve \rho_\xi) \\
& + \frac{\beta_0^4 R}{(1-pq)^2} (2\rho_\xi + (n-2)\rho_\xi^2)\biggr] ,\\
\tilde{V}_3 & \equiv \sum\limits_{j_1\ne k_1; j_2\ne k_2} V^{j_1,k_1; j_2,k_2} = \tilde{V}_0 - \tilde{V}_1 - 2\tilde{V}_2,  \\
\tilde{V}_0 & \equiv \sum\limits_{j_1,k_1, j_2,k_2} V^{j_1,k_1; j_2,k_2} = \sum\limits_{j_1,k_1, j_2,k_2}
\biggl[\C_\ve^{j_1,j_2} \C_\ve^{k_1,k_2} + \frac{2\beta_0^2}{1-pq} \C_\ve^{j_1,j_2} \C_\xi^{k_1,k_2} \\
& + \frac{\beta_0^4 R}{(1-pq)^2} \C_\xi^{j_1,j_2} \C_\xi^{k_1,k_2}\biggr]   
 = n^2 \biggl[(1 + (n-1)\rho_\ve)^2 \\
& + \frac{2\beta_0^2}{1-pq} (1 + (n-1)\rho_\ve) (1 + (n-1)\rho_\xi)  
+ \frac{\beta_0^4 R}{(1-pq)^2} (1 + (n-1)\rho_\xi)^2\biggr], \\
\end{split}
\end{equation}
and we used the identity
\begin{equation}
\C_\ve \C_\xi = \left(\begin{array}{c c c c c} 
 a & b & b & ... & b \\
 b & a & b & ... & b \\
 b & b & a & ... & b \\
...&...&...& ... &...\\
 b & b & b & ... & a \\  \end{array} \right),  \quad 
\begin{cases}  a = 1+(n-1)\rho_\ve\rho_\xi, \\   b = \rho_\ve+\rho_\xi+(n-2)\rho_\ve\rho_\xi  . \end{cases}
\end{equation}
We get then
\begin{equation}
\label{eq:Smulti}
\Sh^2 = \frac{q^2(1-p^2) n (\omega_{11} + \omega_{12}(n-1)\rho_\xi)^2}{\omega_{11}^2 V_1 + 2\omega_{11}\omega_{12} V_2 + \omega_{12}^2 V_3} ,
\end{equation}
where the new coefficients $V_j \equiv \tilde{V}_j
\frac{(1-pq)^2}{n\beta_0^4}$ are obtained from Eqs. (\ref{eq:Vj}).  
Setting to zero the derivative of $\Sh^2$ with respect to
$\omega_{11}$, one gets the optimal weights ratio $x_{\rm opt}$ in
Eq. (\ref{eq:xopt_multi}), as well as the corresponding squared
optimal Sharpe ratio in Eq. (\ref{eq:Sopt_multi}).

Since the explicit formulas for $x_{\rm opt}$ and $\Sh_{\rm opt}$ are
too cumbersome, it is instructive to consider their asymptotic
behavior as $n$ goes to infinity.  Since each asset has $n-1$ lead-lag
terms, it is convenient to introduce the small parameter as $1/(n-1)$
(instead of $1/n$).  In particular, one gets
\begin{equation}
x_{\rm opt} \simeq \frac{-1}{n-1} + \frac{\rho_\xi [Q^2(1-\rho_\ve)^2 + 2Q(1-\rho_\ve)(1-\rho_\xi) + R(1-\rho_\xi)^2]}
{(1-\rho_\xi)[Q^2\rho_\ve^2 + 2Q\rho_\ve \rho_\xi + R\rho_\xi^2]~(n-1)^2} + \ldots 
\end{equation}

\end{document}